\documentclass[reprint]{revtex4-2}
\usepackage{float}
\usepackage{subcaption}
\usepackage{amsmath}   
\usepackage{revsymb}
\usepackage{graphicx}
\usepackage{amsfonts}
\usepackage{bm}
\usepackage{hyperref}
\usepackage{tabularx}

\hypersetup{colorlinks=true, allcolors=blue}
\usepackage[normalem]{ulem}

\newcommand{\fsim}{ \widehat{\textup{fSim}}}

\newcommand{\beq}{\begin{equation}}
\newcommand{\eeq}{\end{equation}}

\newcommand{\adag}{\hat{a}^\dagger}
\newcommand{\ah}{\hat{a}}
\newcommand{\fdag}{\hat{f}^\dagger}
\newcommand{\fh}{\hat{f}}

\begin{document}


\title{Concurrent Fermionic Simulation Gate}

\author{Zhongyi Jiang}
\affiliation{ Peter Gr\"unberg Institute (PGI-2), Forschungszentrum Jülich, Jülich 52428, Germany}
\affiliation{ Institute for Quantum Information, RWTH Aachen University, D-52056 Aachen, Germany}

\author{Mohammad H. Ansari}
\affiliation{ Peter Gr\"unberg Institute (PGI-2), Forschungszentrum Jülich, Jülich 52428, Germany}
\affiliation{ Institute for Quantum Information, RWTH Aachen University, D-52056 Aachen, Germany}

\date{\today}

\begin{abstract}
Introducing flexible native entanglement gates can significantly reduce circuit complexity. We propose a novel gate integrating iSWAP and CPHASE operations within a single gate cycle. We theoretically show one possible realization of this gate for superconducting qubits using bichromatic parametric drives at distinct frequencies. We show how various parameters, such as drive amplitudes and frequencies, can control entanglement parameters.  This approach enhances gate versatility, opening pathways for more efficient quantum computing.
\end{abstract}

\maketitle


\section{Introduction}
\label{sec. introduction}

Quantum computing has made remarkable progress, achieving milestones such as quantum error correction, which enables practical computation despite challenges like decoherence \cite{gottesman1997,acharya2024quantum,bluvstein2024logical}. Modern quantum processors now support around 100 qubits, and quantum supremacy demonstrations have shown limited speedups \cite{arute2019quantum,Acharya:2023aa}. As highlighted recently in Ref. \cite{mohseni2024}, the next goal is achieving quantum utility by efficiently solving practical problems. This requires significant advancements in the challenges for 100-1M physical qubit processors, such as qubit coherence, gate accuracy and compactification \cite{mohseni2024,Bravyi_2022}. 

A quantum processor is considered universal if it can simulate any quantum circuit to arbitrary accuracy using a finite set of gates acting on a finite number of qubits. This requires single-qubit Hamiltonians to generate all $SU(2)$ gates and at least one two-qubit interaction to produce entangling gates \cite{Nielsen_Chuang_2010}. Discussions of universality often focus on unitary gates rather than physical Hamiltonians.

Recent technological advancements for controlling non-unitary errors, e.g. nonequilibrium quasiparticle tunneling \cite{mcewen2024,Benevides_2024,PhysRevB.91.195434,Ansari_2015} and trap state noise\cite{Lisenfeld:2023aa,Ansari_2013},  as well as unitary errors, e.g. $ZZ$ stray couplings \cite{Xu2024,Xu2021,PhysRevApplied.19.024057,fors2024comprehensive,PhysRevLett.107.080502,Kandala_2021,PhysRevLett.129.040502,Mitchell_2021,PhysRevLett.129.060501,Ding_2023,PhysRevLett.127.080505,PhysRevLett.125.240502}, have improved gate fidelity, architecture, and manufacturing, have enabled problem-specific algorithms to leverage quantum computing’s power \cite{Preskill2018quantumcomputingin,Acharya:2023aa,Egan2021FaulttolerantCO,RevModPhys.86.153,Daley2022PracticalQA,bauer2022quantum,NatureEvidenceUtility}. These developments have set the stage for high-fidelity computing with thousands of qubits. Two of the most commonly used two-qubit gates are iSWAP  (swapping $|01\rangle \leftrightarrow |10\rangle$) and CPHASE (applying a $\pi$ phase to $|11\rangle$). Often, a small CPHASE is a byproduct of iSWAP and vice versa. However, it has been shown recently that one can better control the CPHASE byproduct of iSWAP  to define a new gate, namely `fermionic Simulation (fSim).' This new gate can be separated into two gates in certain conditions. 

The fSim gate is a versatile quantum computing tool, particularly useful for simulating phase transitions in strongly correlated fermionic systems \cite{PhysRevA.79.032316,PhysRevLett.120.110501,PhysRevB.102.235122,Verstraete2009}. Combining the functionality of two gates reduces resource demands and enhances flexibility for parameter-specific applications.

The fSim gate has been implemented in superconducting qubits using bichromatic modulation of frequency detuning and coupling strengths or a DC pulse with a parametric drive \cite{PhysRevLett.125.120504,npjfSimStaticAndPara}. These approaches enable continuous transformations, swapping $|01\rangle$ and $|10\rangle$ while introducing a controllable phase for $|11\rangle$. The gate’s parameters $\theta$ and $\varphi$ allow state swaps and apply a phase factor $\exp(i\varphi)$, as represented by:

\begin{equation}\label{eq.pmatrix}
    \fsim(\theta,\varphi)=\begin{pmatrix}
  1 & 0 & 0 & 0\\ 
  0 & \cos(\theta) & -i\sin(\theta) & 0\\
  0 & -i\sin(\theta) & \cos(\theta) & 0\\
  0 & 0 & 0 & e^{i\varphi}
  \end{pmatrix}
\end{equation}

With full control over $\theta$ (0 to $90^\circ$) and $\varphi$ (0 to $180^\circ$), the fSim gate mimics anti-commuting particles. It improves the Quantum Approximate Optimization Algorithm (QAOA) \cite{PRXQuantum.1.020304} and reduces circuit depth in quantum variational eigensolvers for molecular hydrogen by a factor of 10 compared to controlled-NOT gates \cite{PhysRevApplied.11.044092}. The rotation $\fsim(\pi/2, \pi)$ simplifies resource use for fermion simulations. Combining native fSim gates with {XYZ} decomposition and optimized fermion-to-qubit mappings reduces circuit depth by 70\% in Fermi-Hubbard model simulations on square lattice quantum processors \cite{Algaba2024lowdepthsimulations}.

As an example, consider the general fermionic Hamiltonian: $H=\sum_{ij}A_{ij} \fdag_i \fh_j + \sum_i B_i \hat{n}_i + \sum_{i\neq j} C_{ij} \hat{n}_i \hat{n}_j$, with $\hat{n}_i=\fdag_i \fh_i$ as the number operator. This Hamiltonian appears in models like the Hubbard model and quantum chemistry, with hopping terms $A_{ij}$, on-site energy $B_i$, and electron-electron interaction $C_{ij}$. In quantum simulation, the on-site energy is mapped to single-qubit gates. The hopping terms correspond to iSWAP operations. The electron-electron interaction together with fermionic anti-commutator can be mapped to the conditional phase $\varphi$. Therefore, being able to implement iSWAP and CPHASE operations simultaneously makes simulating such kind of Hamiltonians more efficient.

Although the fSim concept offers several advantages for quantum algorithm design, some challenges
hampered previous fSim implementations. In experiments, calibrating the gate in a continuous 2D parameter space is very time-consuming. 
From the theory side, there is no well-established theory for driving iSWAP transition and CPHASE transition simultaneously. One of the reasons is that for systems where two-qubit gates are activated via level detunings and static couplings, detunings and couplings in the one-excitation subspace (between $|01\rangle$ and $|10\rangle$) and two-excitation subspace (among $|11\rangle$, $|02\rangle$ and $|20\rangle$) can not be controlled independently, which makes the iSWAP transition and CPHASE transition correlated\cite{PhysRevLett.125.120504}. For systems where two-qubit gates are driven via fast oscillating pulses ( microwave drives or parametric flux drives), the lack of understanding of drive-drive crosstalk makes it difficult to control simultaneous transitions.

To address those issues in theory, we propose a theoretical modification to the fSim gate that integrates iSWAP and CPHASE operations into a single hardware. This enables their simultaneous and controllable application, namely the “concurrent fSim (cfSim).” This approach compresses the gate time of consecutive iSWAP and CPHASE operations into one shorter gate with the same output.  Our theoretical study demonstrates that this method simultaneously achieves the full range of iSWAP angles and conditional phases, inherently facilitating a fermion swap in a single operation.  We note that there have been several previous works in recent
years on concurrent driving of gates to implement complicated
gates\cite{gu2021fast,kim2022high,warren2023extensive,PhysRevApplied.21.034018}.  In fact, in a recent experiment, it is suggested to run iSWAP and CZ simultaneously through a bichromatic parametric drive on the coupler between two qubits\cite{krivzan2024quantum}. In this paper, we give a detailed analysis and simulations of how concurrent fSim gates can be implemented. We employ the example of bichromatic parametric drives at distinct frequencies in a transmon-coupler-transmon setup.  The method uses two parametric drives: one resonating with the $|01\rangle \leftrightarrow |10\rangle$ transition to control the iSWAP angle, and the other facilitating CPHASE transitions via $|11\rangle \leftrightarrow |02\rangle$, $|11\rangle \leftrightarrow |20\rangle$, or both. Completing an oscillation cycle induces a detuning-dependent phase shift in the $|11\rangle$ state and reduces leakage at the same time. Moreover, we develop precise analytical strategies to accurately predict transition dynamics and resolve crosstalk between the drives, resolving synchronization issues. 

The remainder of the paper is organized as follows. In Sec. \ref{sec. model}, we introduce the setup of bichromatic parametric drive and study two models based on it to understand the dynamics of two-tone drives. We first present analytical derivations of a two-qutrit toy model and then extend it to the more complete coupled Kerr nonlinear oscillator model. Section \ref{sec.Opt} deals with an important aspect of superconducting quantum gates: leakage. We show how to minimize leakage when two drives are present at the same time. In Section \ref{sec:cfsim}, we combine all analyses above to derive analytical expressions of iSWAP angle $\theta$ and conditional phase $\varphi$ in cfSim gates. Numerical simulation and optimization of cfSim are performed. Sec. \ref{sec.ZZ free} presents an important case of cfSim gates: ZZ-free iSWAP gate. In Sec.\ref{sec.Discussion and outlook}, we discuss what limits the cfSim gate fidelity and give some outlook of our proposed gate scheme. In Section \ref{sec.Summ}, we summarize this paper.

\section{The Model}
\label{sec. model}

In superconducting circuits consisting of two transmons coupled via a frequency-tunable coupler, a continuous two-qubit fSim gate has recently been realized. iSWAP and CPHASE gates can be implemented by adjusting the qubit-qubit coupling strength and frequency detuning, with one gate activated immediately after the other is deactivated \cite{PhysRevLett.125.120504}. Notably, both gates can also be active simultaneously, enabling the complete and concurrent implementation of the fermionic simulation gate set in a single operation, referred to as concurrent fermionic simulation (cfSim). Below, we present the theory for realizing such a gate using two parametric drives in a controllable manner.

\subsection{Setup: bichromatic parametric driving (BPD)}
\label{subsec:setup BPD}

\begin{figure}[ht]
\includegraphics[width=0.45\textwidth]{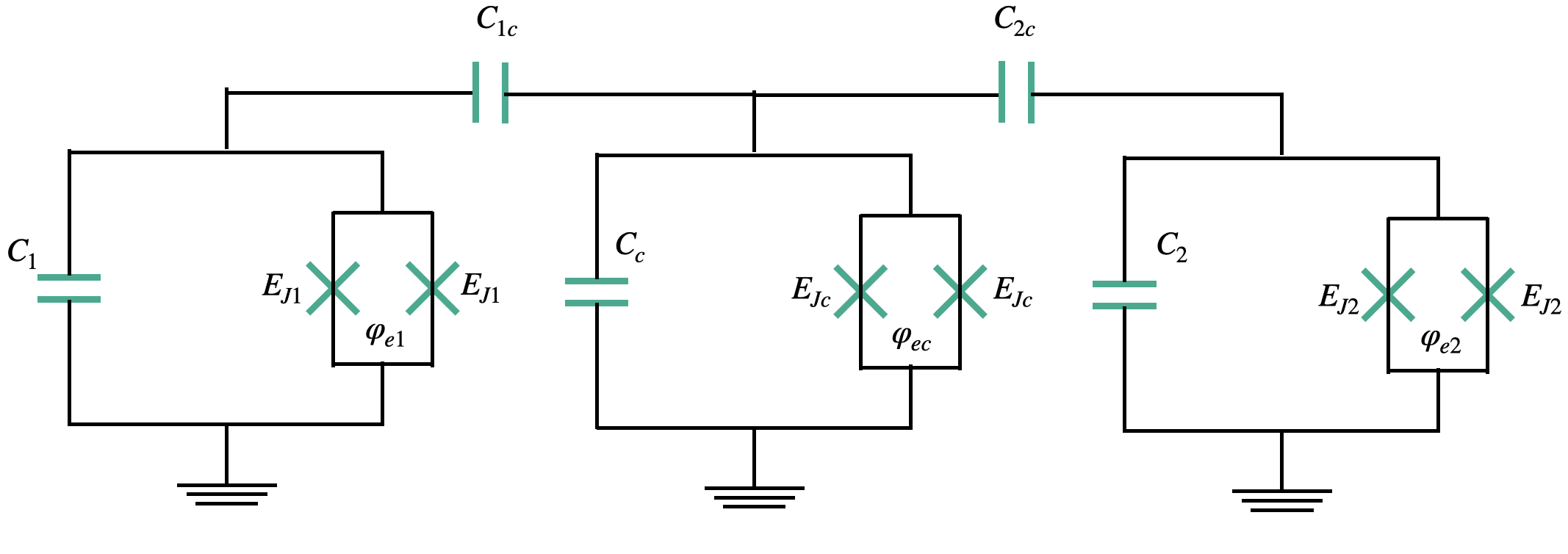}
\caption{Setup of the system. Two transmons are coupled via a common coupler. The direct transmon-transmon coupling is ignored. Two parametric drives are applied to the two transmons separately.  The two simultaneous drives induce a concurrent fSim gate between the two qubits, exchanging their excitations and adding a conditional phase.}
\label{fig:setup}
\end{figure}

Motivated by the idea of realizing fSim gate by a parametric drive \cite{Reagor_2018,caldwell2018parametrically,PRXQuantum.5.020325}, in this paper, we consider the simultaneous application of two gates achieved by employing Bichromatic Parametric Driving (BPD). In this approach, we apply two parametric drives at distinct frequencies, each on one qubit. Our proposed gate operates by modulating the frequencies $\omega_1$ and $\omega_2$ in an oscillatory manner. Consider the setup depicted in Fig.(\ref{fig:setup}), where both transmons $Q_1$ and $Q_2$, coupled via a resonator, are parametrically driven via modulating external phases $\varphi_{e1}$ and $\varphi_{e2}$.

The full circuit Hamiltonian reads\cite{petrescu2023accurate}:

\begin{equation}\label{eq:full circuit H}
    H= H_{1}+H_{2}+H_c+H_{int}
\end{equation}
where the transmon Hamiltonians, coupler Hamiltonian, and the coupling Hamiltonian are expressed in the canonical phase-charge basis as:

\begin{equation}\label{eq:full circuit H in n p}
\begin{split}
    H_j&=4E_{Cj} \mathbf{ \hat{n}}_j^2-E_{Jj} \cos( \bm{\hat{\varphi}}_j+\varphi_{ej}(t))\\ 
    H_{int}&=\Sigma_{jk} 4E_{Cjk} \bm{\hat{n}}_j \bm{\hat{n}}_k
\end{split}
\end{equation}
with $j,k = 1,2  \text{ and 
 } c$.

The analytical analysis and numerical simulation of the full circuit Hamiltonian is quite complicated. To focus on the essential points of our gate scheme, we instead study two simplified models. We first introduce a toy model consisting of two interacting qutrits to understand the basics of the gate dynamics under BPD. We then turn to a more complete model including transmon higher levels and coupler degrees of freedom, which is more accurate yet can still be studied analytically. Exact numerical simulation of the full circuit model can be performed with the help of methods such as time-dependent Schrieffer-Wolff transformation or Floquet theory\cite{petrescu2023accurate}.

\subsection{A Toy Model: Two Qutrits}
\label{subsec. weakbpd}

\begin{figure}[t]
\includegraphics[width=0.45\textwidth]{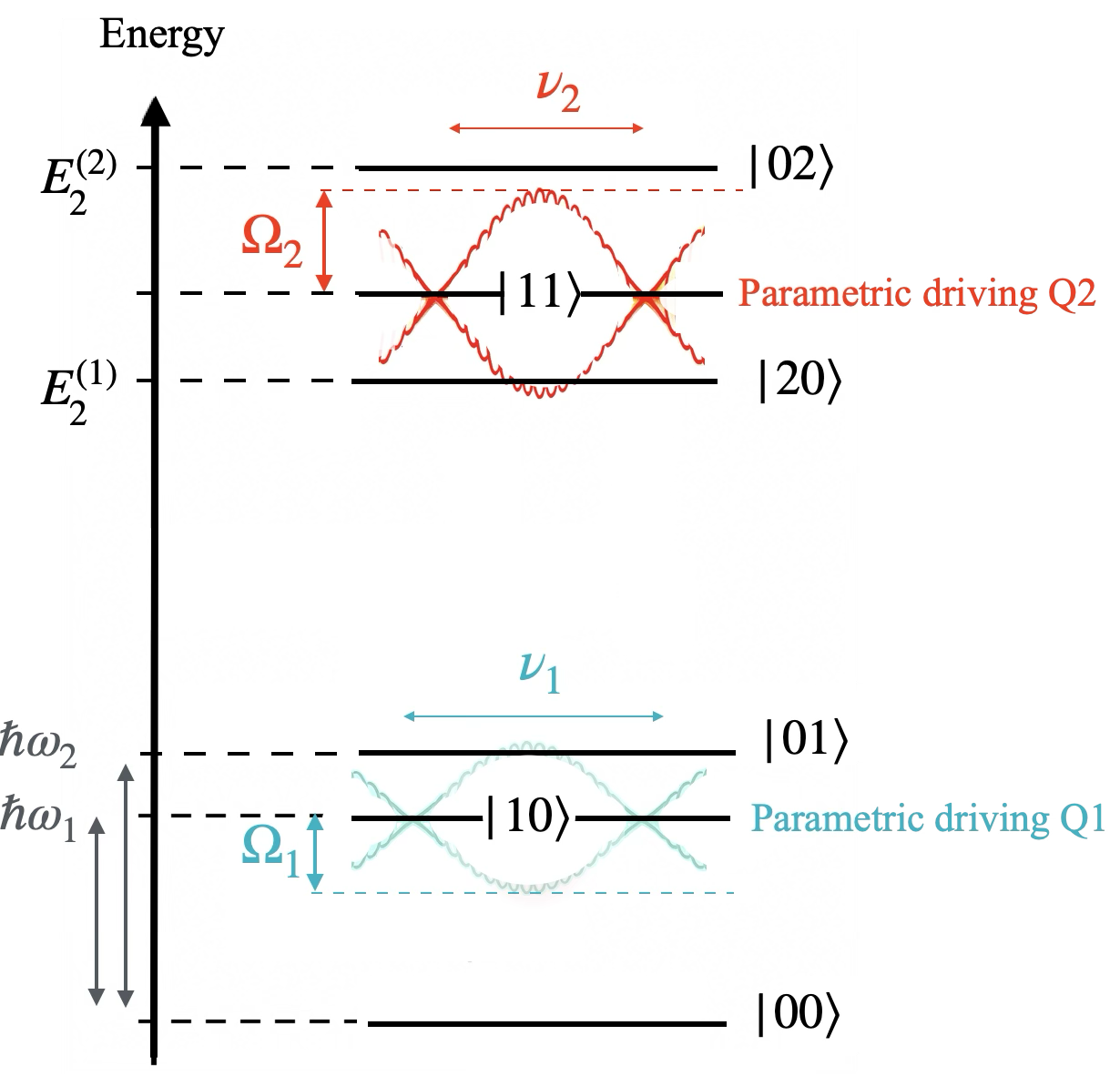}
\caption{{Schematic of BPD for cfSim gates. Relevant levels are shown. Two parametric drives target two transitions separately. When on resonance, drive frequency $\nu_{1,2}$ should match level detuning. Drive amplitudes $\Omega_{1,2}$ are not required to match detunings. For cfSim gates, we choose to drive 1 to resonantly drive $|01\rangle \leftrightarrow |10\rangle$ transition, therefore, $\nu_1=\omega_2-\omega_1$. The second drive is chosen to be near resonant with $|11\rangle \leftrightarrow|20\rangle$ or $|11\rangle \leftrightarrow |02\rangle$ transition.} iSWAP and CPHASE transitions are happening simultaneously. Both iSWAP angle $\theta$ and conditional phase $\varphi$ can be controlled.}
\label{fig:simultaneous iswap  and cz schemetics}
\end{figure}

The schematic of the proposed gate scheme is depicted in Fig.~(\ref{fig:simultaneous iswap  and cz schemetics}). The first drive resonates the state $|11\rangle$ with either $|02\rangle$ or $|20\rangle$. The second drive facilitates a swap between the single-qubit excited states $|10\rangle$ and $|01\rangle$ by bringing them into resonance, as illustrated in Fig.(\ref{fig:simultaneous iswap  and cz schemetics}). The state $|00\rangle$ remains uncoupled from other states because it is far detuned from the relevant transitions, thereby staying unaffected by the applied drives.

It is crucial to minimize leakage out of the computational space, which requires the state $|11\rangle$ to complete full oscillation cycles, transferring its population out of the computational subspace (to $|02\rangle$ or $|20\rangle$) and then back. This depopulation and repopulation of $|11\rangle$ are optimally synchronized with the swap transition between $|01\rangle$ and $|10\rangle$.

The qubit dynamics of an iSWAP /CPHASE gate set can be efficiently modeled by two interacting qutrits with time-dependent frequency detuning $\omega_1(t)-\omega_2(t)=\Delta + h_1(t)-h_2(t)$ and fixed coupling strength $g$. The static Hamiltonian of such a  system in the basis of  $|01\rangle$, $|10\rangle$, $|11\rangle$, $|02\rangle$, $|20\rangle$ from left (top) to right (bottom) is given by:
\beq
\label{eq.Hqutrits}
    H_0=\begin{pmatrix}
  0 & g & 0 & 0 & 0 \\ 
  g & \Delta & 0 & 0 & 0  \\
  0 & 0 & \Delta & \sqrt{2} g & \sqrt{2} g \\
  0 & 0 & \sqrt{2} g &  \delta & 0 \\
  0 & 0 & \sqrt{2} g & 0 & 2\Delta+\delta 
  \end{pmatrix}
\eeq
with similar qutrit anharmonicity $\delta_1=\delta_2\equiv \delta$ with $\delta_i \equiv E^{(i)}_2-2 \hbar \omega_i$ denoting the anharmonicty associated to qutrit $i$.

Since a parametric drive modifies the qubit frequency, it effectively acts as a time-dependent number operator $\adag_i \ah_i$, where $\adag_i$ ($\ah_i$) represents the creation (annihilation) operator for the qutrit $i$.  For simplicity, we assume anharmonicities are not affected by drives. The total Hamiltonian within the $5 \times 5$ qutrit Hilbert space of interest is expressed as\cite{PhysRevLett.125.120504,Reagor_2018,caldwell2018parametrically}:

\beq
\label{eq:H full}
    H=H_0+\begin{pmatrix}
  h_2(t) & 0 & 0 & 0 & 0 \\ 
  0 & h_1(t) & 0 & 0 & 0  \\
  0 & 0 & \sum_{i}h_i(t) & 0 & 0 \\
  0 & 0 & 0 & 2 h_2(t) & 0 \\
  0 & 0 & 0 & 0 & 2 h_1(t) 
  \end{pmatrix}
\eeq

Let's assume the drives $h_1$ and $h_2$ take the form: $h_i=\Omega_i \sin (\nu_i t)$, where drive amplitudes and frequencies are time-independent. We can then move to a rotating frame where all diagonal terms are zero, see Appendix (\ref{Appen: rotating frame}). In this rotating frame, the Hamiltonian is given by:

\begin{equation}\label{eq. HBessel}
    \begin{split}
      H  = &\sum_{m,n=0}^\infty  gJ_m \left(\frac{\Omega_1}{\nu_1}\right) J_n \left(\frac{\Omega_2}{\nu_2}\right)  e^{-i t \left(m\nu_1+n\nu_2\right)} \times\\  
      & \qquad  \qquad  ( e^{i t \Delta} |10\rangle \langle 01| + \sqrt{2} e^{i t (\delta+\Delta)} |11\rangle \langle 02|   \\  &\qquad  \qquad  \qquad + \sqrt{2}e^{i t (\delta-\Delta)} |11\rangle \langle 20|  +H.c.)
   \end{split}
\end{equation}
where $\Delta_0$ is the static detuning between $|01\rangle$ and $|10\rangle$, $J_n$ is the Bessel function of the first kind, and the summation takes place over integers $n\in \mathbb{Z}$. 

In Appendix (\ref{Appen:perutbation of toy model}), we show that, under certain perturbative conditions,  the Hamiltonian can be further simplified:
\begin{eqnarray}
\label{eq. HBessel_simplf}
      H  &=& gJ_1(x_1) J_0(x_2) |10\rangle \langle 01| \nonumber \\ && + \sqrt{2} gJ_0(x_1) J_1(x_2) e^{i\varepsilon t}|11\rangle \langle 02| + h.c.
\end{eqnarray}
with $x_k=\frac{\nu_k}{\Omega_k}$.

The residual time dependence in the coupling between $|11\rangle $ and $ |02\rangle$ can be treated by applying another rotating frame transformation defined as $U_R=e^{-i\varepsilon t |02\rangle \langle 02|} \otimes I$, which rotates $|02\rangle$ state and leaves other levels unchanged. This helps to transform the interaction to the frame rotating with the frequency of level detuning between $|11\rangle$ and $|02\rangle$.  The resulting Hamiltonian turns out to be block-diagonal with the following two diagonal blocks,  that resemble separate iSWAP gates in the subspace spanned by $|10\rangle, |01\rangle$  and CPHASE gate in the subspace spanned by $|11\rangle, |02\rangle$, as follows:
\begin{eqnarray}
\label{eq. Happrox}
  H_{10,01}&=\begin{pmatrix}
  0 & gJ_1(x_1) J_0(x_2) \\ 
 gJ_1(x_1) J_0(x_2) & 0 \\
  \end{pmatrix}  \\
 H_{11,02} &= \begin{pmatrix}
  0 & \sqrt{2}g J_0(x_1) J_1(x_2) \\ 
  \sqrt{2}g J_0(x_1) J_1(x_2) & -\varepsilon \\
  \end{pmatrix} 
\end{eqnarray}

Note that in our numerical analysis, we carry on the analysis in the presence of all possible transitions, and therefore, we consider the weak driving limit of this section only to make analysis simpler for analytical verification. Performing the analytical study in the relatively intermediate regime of driving amplitude requires to reconsider the coupling $|11\rangle \langle 20 |$ that needs to deal with $3 \times 3$ matrix size everywhere.

We have now established that it is possible to express parametrically driven CPHASE and iSWAP operations using a 2x2 matrix representation. Consequently, we propose describing the operators for the iSWAP and CPHASE gates using an identical matrix. To facilitate this approach, we can mathematically introduce the following hypothetical Hamiltonian:
\begin{equation}\label{eq:tls matrix}
  H=\begin{pmatrix}
  0 & g \\ 
  g & \Delta \\
  \end{pmatrix}
\end{equation}

In the case of iSWAP the two columns (and rows) indicate the states $|01\rangle$ and $|10\rangle$ with $g$ being the interchanging strength between the two, which is nearly the interaction strength between the two qubits in the computational level, $\Delta$ the frequency detuning of the two qubits. Yet the very matrix of Eq.~(\ref{eq:tls matrix}) can represent the CPHASE operator once the columns (and rows) are labelled by $|11\rangle$ and $|02\rangle$. Therefore   $g$ for the CPHASE operator will be the interaction strength between the highest excited level in the computational subspace of two qubits with the closest non-computational energy level.

Now consider that the state of two isolated qubits is evolved for a specific time $t$ only by the unitary operator $U(t)$, i.e., $U(t)=e^{-iHt}$. In the two-level picture we chose for the two gates in Eq.~(\ref{eq:tls matrix}), the unitary state evolution operator can be represented as  $U(t)=e^{-i t \Delta /2} \left(\cos(\Omega t) \hat{1}-\ i \sin(\Omega t) \left(n_z\hat{Z}+n_x \hat{X}\right)\right)$, with $\hat{X}$ and $\hat{Z}$ being the Pauli matrices, the Rabi frequency $\Omega= \sqrt{(\Delta/2)^2+g^2}$ determining the rotation rate. The Pauli coefficients $n_z=-\Delta/2\Omega $ and $n_x=g/\Omega$ define the rotation axes.

The iSWAP transition usually occurs by bringing the frequency of one qubit in resonance with the other qubit or applying a resonant drive with the qubit-qubit detuning.  In the case of applying a resonant drive, the frequency detuning $\Delta$ between the two qubits becomes zero in the rotating frame. In this case one can show $| \langle 01|U(t)|10\rangle |= \sin (g t)$ which indicates the rotation iSWAP  angle $\theta$ can be obtained by applying the gate $U(t)$ for certain time; i.e.   $\theta\equiv \arcsin (| \langle 01|U(t)|10\rangle |)=gt$. 

The CPHASE gate requires frequency detuning $\Delta$ between the two qubits. In this gate transition sends the photon outside of the computational level, however, in order to minimize leakage, it is demanded that the final state goes back to  $|11\rangle$ at the end of the cycle of two CPHASE evolutions. Therefore the total evolution adds a phase to the state $|11\rangle$. 
\begin{equation}
    U(t_g)=-e^{-i\frac{\Delta t_g}{2}}
\end{equation}
with $t_g=\frac{2\pi}{\sqrt{\Delta^2+4g^2}}$ 

So, the conditional phase $\varphi$ is given by 
\begin{equation}\label{eq:tunable phi}
\varphi=\pi(1-\frac{\Delta }{\sqrt{\Delta^2+4g^2}})    
\end{equation}

In terms of $t_g$ it can be simplified as:

\begin{equation}\label{eq:tunable phi as tg}
\varphi=\pi(1-\frac{\Delta }{2\pi}t_g)    
\end{equation}

Therefore, the iSWAP angle $\theta$ and the conditional phase $\varphi$ can be controlled by tuning the corresponding coupling strengths and drive detuning.

This is not the only way to drive a tunable conditional phase. In another approach, one drives the $|10\rangle \leftrightarrow |02\rangle$ transition with zero detuning and stays at $|02\rangle$ for varying time to accumulate different conditional phases\cite{scarato2025realizing}.  The accumulated phase is given by $\varphi= (\omega_{20}-\omega_{10}-\omega_{01})\tau$, with $\tau$ being the waiting time at $|20\rangle$. This gate can be implemented fast because $\omega_{20}-\omega_{10}-\omega_{01}$ is usually in the hundred MHz range. The total conditional phase combines the dynamical and geometric phases in the above cases. It is also possible to have a pure geometric conditional phase by properly choosing the evolution trajectory on the Bloch sphere\cite{li2021highGeometricphase}.

\subsection{Full Model: Transmon-Coupler-Transmon}
\label{subsec.strongbpd}

Let us once again consider the case of two flux-tunable transmons coupled via a harmonic resonator, as depicted in the hardware schematic of Fig.~(\ref{fig:setup}). Here we consider a larger Hilbert space for each transmon and coupler. Bichromatic parametric drives are applied to the device, with each drive acting on one transmon. For simplicity of analysis, we consider the setup where one drive is applied to each transmon individually. However, this is not a necessary condition from a theoretical perspective. Drives could instead be applied to the resonator, if it is also tunable, or both drives could be applied to a single transmon or the coupler\cite{valery2022dynamical}.

As discussed at the beginning of this section, treating the full circuit Hamiltonian as in Eq.~(\ref{eq:full circuit H}) and Eq.~(\ref{eq:full circuit H in n p}) is not an easy task, of which the complexity might obscure the physics behind BPD dynamics. We therefore decide to model the circuit as two Kerr nonlinear oscillators coupled via a harmonic oscillator. We furthermore assume that the drives only modulate the frequencies of the oscillators but not their anharmonicities. With those approximations, the dominant features of the physics are kept, and analytical formulas are still feasible.

Neglecting the counter-rotating terms, we can write down the Hamiltonian of the system:

\begin{equation}\label{eq:setup}
    \begin{split}
    &  H=H_0+H_d,\\ 
        & H_0=\sum_{\alpha=1,2,c} \omega_\alpha \adag_\alpha \ah_\alpha + \frac{\delta_j}{2} \adag_\alpha \ah_\alpha (\adag_\alpha \ah_\alpha-1) \\ & \quad \quad \quad + \sum_{j,k=\{1,2\}} g_{jc} (\adag_j \hat{a}_c+ H.c.),\\
        & H_d=\sum_{j=1,2} \Omega_j \sin(\nu_j t + \phi^d_j) \adag_j \ah_j.
    \end{split}
\end{equation}
where the subindex $c$ denotes the coupler, $g_{jc}$ is the coupling strength between transmon $j$ and the coupler. For the case of a harmonic coupler one expects that $\delta_c=0$, but in general the coupler can also have a non-zero anharmonicity.

The total Hamiltonian $H$ in Eq.~(\ref{eq:setup}) consists of the undriven Hamiltonian $H_0$ and the drive Hamiltonian $H_d$. The undriven Hamiltonian $H_0$ is the generalized Jaynes-Cummings.
Hamiltonian\cite{PhysRevA.76.042319}. The drive Hamiltonian has the standard form of parametric drives\cite{{roth2017analysis},{PhysRevB.73.064512},{niskanen2007quantum},{PhysRevB.73.094506},{PhysRevA.86.022305},{PhysRevB.87.220505},{royer2017fast},{naik2017random},{PhysRevApplied.6.064007},{caldwell2018parametrically},{PhysRevResearch.2.033447}}.

In order to accurately study the dynamics of the system, we first find a transformation $U_0$ that  diagonalizes $H_0$, i.e. $\Tilde{H}_0=U_0^\dag H_0 U_0$ with $\Tilde{H}_0$ denoting diagonalized $H_0$. By transforming the drive Hamiltonian to the same diagonal frame where in $H_0$ is diagonalized, the drive Hamiltonian becomes $\Tilde{H}_d=U_0^\dag H_d U_0$, which is not necessarily a diagonal matrix. The total Hamiltonian in the diagonal frame is then $\Tilde{H}_0 + \Tilde{H}_d$. Note that in this treatment, we are studying the dynamics in the dressed basis of the undriven Hamiltonian. Note that the diagonal transformation $U_0$ can be analytically found  using the Bogoliubov transformation\cite{{RevModPhys.93.025005},{PhysRevA.98.053808},{PhysRevB.100.024509}} or by exact numerical diagonalization. Here, in order to perform precise analysis we numerically determine the diagonalization transformation matrix $U_0$. This does not add too much computational cost when the undriven Hamiltonian $H_0$ is fixed.

Let us denote dressed transmon states by $Q_1$ and $Q_2$, and  dressed coupler(a harmonic resonator in our analysis ) states with $R$. In the Fock basis $\{|Q_1 Q_2 R\rangle\}$. Total Hamiltonian reads:

\begin{widetext}
\begin{equation} \label{eq:total H diag frame}
\begin{split}
\Tilde{H}&=\sum_{Q_1, Q_2, R} \left\{ \left( \Tilde{\omega} _{Q_1Q_2R}+ \sum_{m=1,2} \Omega_m \sin(\nu_m t+\phi^d_m )  N^m_{Q_1Q_2R} \right) |Q_1Q_2R\rangle \langle Q_1Q_2R| \right. \\ 
& \quad \quad \quad  \quad \left. +\sum_{Q_1^\prime, Q_2^\prime, R^\prime} \sum_{m=1,2}  \Omega_m \sin(\nu_m t+\phi^d_m) C^m_{Q_1^\prime Q_2^\prime R^\prime; Q_1 Q_2 R}  |Q_1^\prime Q_2^\prime R^\prime\rangle \langle Q_1 Q_2 R|  \right\} 
\end{split}
\end{equation}
\end{widetext}
with $\Tilde{\omega}_{Q_1 Q_2 R}$,  $N^m_{Q_1 Q_2 R}$ and $C^m_{Q_1^\prime Q_2^\prime R^\prime; Q_1 Q_2 R}$ being defined as follows:
\begin{equation}\label{eq:N C} \begin{split}
        & \Tilde{\omega}_{Q_1 Q_2 R}=\langle Q_1 Q_2 R |\Tilde{H}_0| Q_1 Q_2 R\rangle \\
        & N^m_{Q_1 Q_2 R}=\langle Q_1 Q_2 R |U^\dag_0 \adag_m \ah_m U_0 |  Q_1 Q_2 R\rangle \\
        & C^m_{Q_1^\prime Q_2^\prime R^\prime; Q_1 Q_2 R}=\langle Q_1^\prime Q_2^\prime R^\prime |U^\dag_0 \adag_m \ah_m U_0 |  Q_1 Q_2 R\rangle
    \end{split} \end{equation}
defining $\Tilde{\omega}_{Q_1 Q_2 R}$ as the eigenfrequency of the dressed state $|Q_1 Q_2 R\rangle$.  The transformed operator $U^\dag_0 \adag_m \ah_m U_0$ is split into the diagonal term $N^m_{Q_1 Q_2 R}$ and the off-diagonal term $C^m_{Q_1^\prime Q_2^\prime R^\prime, Q_1 Q_2 R}$. The two parametric drives of BPD are denoted by $\Omega_m \sin(\nu_m t+\phi^d_m)$ modulates the eigenfrequencies of the dressed state $|Q_1 Q_2 R\rangle$ through the diagonal term $N^m_{Q_1 Q_2 R}$. These drives also introduces time dependent coupling between $|Q_1 Q_2 R\rangle$ and $|Q_1^\prime Q_2^\prime R^\prime\rangle$ via  the off-diagonal term $C^m_{Q_1^\prime Q_2^\prime R^\prime; Q_1 Q_2 R}$.

We define the rotating frame corresponding to the diagonal part of $\Tilde{H}$:
\begin{equation}\label{eq:U_r}
\begin{split}
U_r&=e^{-i \int dt \sum_{Q_1 Q_2 R} |Q_1 Q_2 R\rangle \langle Q_1 Q_2 R |\Tilde{H}|  Q_1 Q_2 R\rangle \langle Q_1 Q_2 R| }\\
&=\sum_{Q_1 Q_2 R} e^{-i \int  \langle Q_1 Q_2 R|\Tilde{H} | Q_1 Q_2 R\rangle dt}|Q_1 Q_2 R\rangle \langle Q_1 Q_2 R|
\end{split}
\end{equation}

We transform the Hamiltonian in to the rotating frame $\Tilde{H}_{R}=U_r^\dag \Tilde{H} U_r-i U_r^\dag\partial_t U_r$ and simplify to find:
\begin{equation}\label{eq:H R}
    \Tilde{H}_{R}=\sum \Tilde{G}(t)_{Q_1 Q_2 R;Q_1^\prime Q_2^\prime R^\prime} |Q_1^\prime Q_2^\prime R^\prime\rangle \langle Q_1 Q_2 R| 
\end{equation}

For the full derivation and the expression of $\Tilde{G}(t)_{Q_1 Q_2 R;Q_1^\prime Q_2^\prime R^\prime}$ see Appendix (\ref{appen:G}).

\subsubsection{Monochromatic parametric driving case}
\label{subsubsec.MPD}

Let us consider the simple and special case of single parametric driving, namely `Monochromatic Parametric Drive (MPD).' We apply MPD to Q1 to evaluate the time-dependency of the coupling strength between the swap between transmons, $|010\rangle \leftrightarrow |100\rangle$.  The drive frequency is resonant with the transition: $\nu_1=|\Tilde{\omega} _{010}-\Tilde{\omega} _{100}|$.

\begin{figure*}[ht]
\centering
\begin{subfigure}{0.35\linewidth}
\includegraphics[width=1\textwidth]{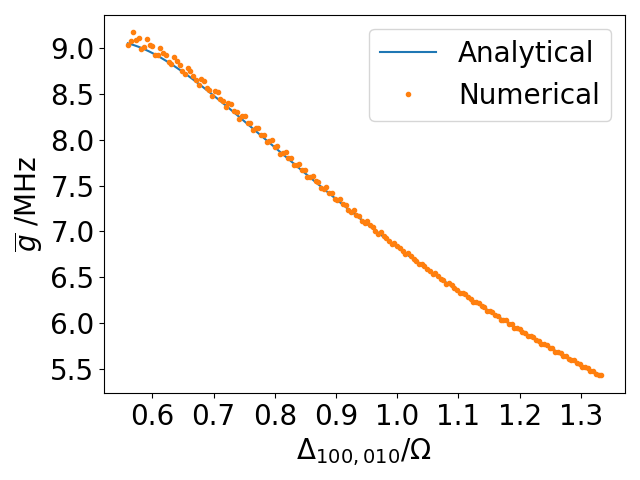}\put(-120,90){\textbf {(a)}}
\end{subfigure}
\begin{subfigure}{0.35\linewidth}
\includegraphics[width=1\textwidth]{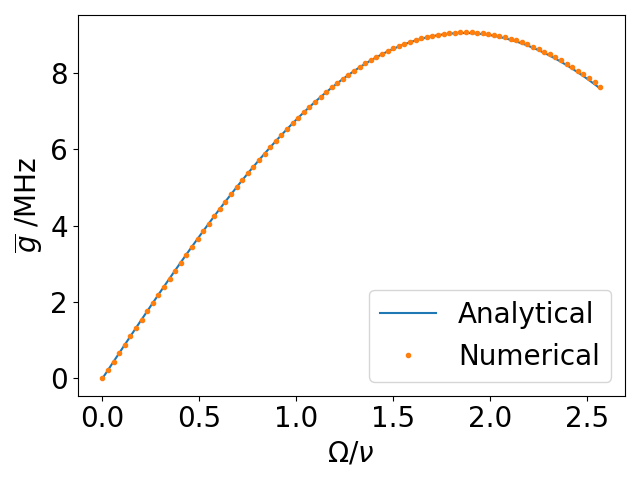}\put(-120,90){\textbf {(b)}} \end{subfigure} 
\begin{subfigure}{0.35\linewidth}
\includegraphics[width=1\textwidth]{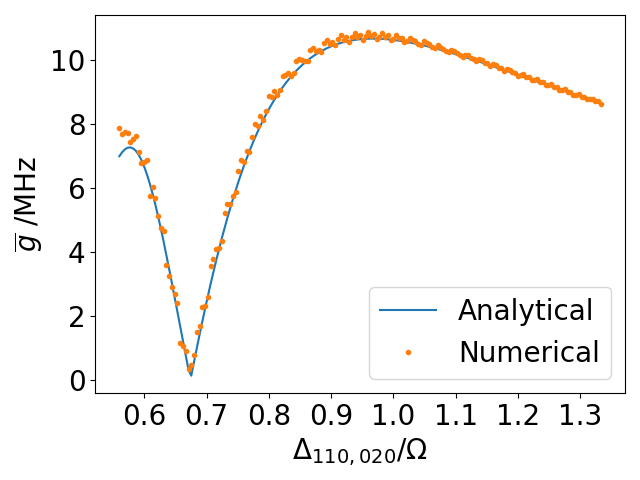}\put(-120,90){\textbf {(c)}}
\end{subfigure}
\begin{subfigure}{0.35\linewidth}
\includegraphics[width=1\textwidth]{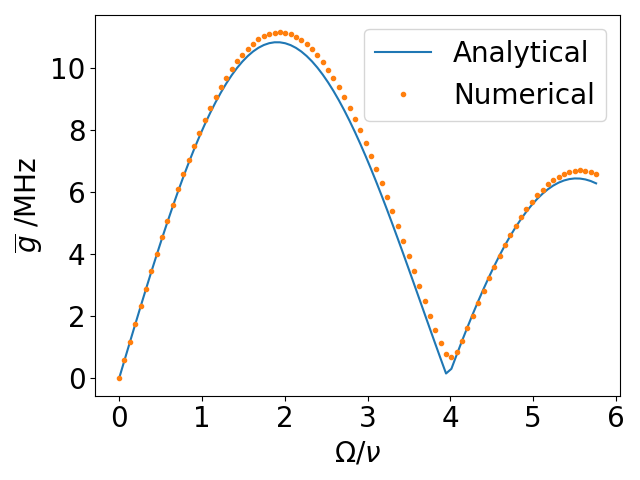}\put(-120,90){\textbf {(d)}}
\end{subfigure}
\caption{Static coupling strength, analytical and numerical. Analytical results are obtained via Eq.~(\ref{eq:g01 10 compact single drive}). Numerical results are from solving time-depenent  Schr\"{o}dinger equation defined by Hamiltonians of Eq.~(\ref{eq:setup}), where the Hilbert space is truncated to 100 dimensions. 5 levels are used for each transmon, and 4 levels are used for the coupler mode. The analytical formula is valid for a wide range, from dispersive regimes to strong coupling regimes. \ref{fig:gs single drive}(a) Static $\bar{g}_{100,010}$ vs qubit detuning. \ref{fig:gs single drive}(b) Static $\bar{g}_{100,010}$ vs drive amplitude. \ref{fig:gs single drive}(c) Static $\bar{g}_{110,020}$ vs qubit detuning.  The dip around 300 MHz corresponds to the root of $J_0(x)+J_2(x)$. \ref{fig:gs single drive}(d) Static $\bar{g}_{110,020}$ vs drive amplitude. The analytical formula can well capture non-linear dependence on $\Omega$.}
\label{fig:gs single drive}
\end{figure*}

In Appendix (\ref{appen:G})  we generalize the case where transmons and coupler are flux-tunable and we apply three parametric drives, two on qubits with frequencies $\nu_1$ and $\nu_2$, and one with frequency $\nu_0$ on the coupler. For this general case we worked out the resonant condition for transition  $|Q_1 Q_2 R\rangle \leftrightarrow |Q_1^\prime Q_2^\prime R^\prime\rangle$ will be $\Tilde{\omega} _{Q_1^\prime Q_2^\prime R^\prime}-\Tilde{\omega} _{Q_1 Q_2 R}-(n_0\nu_0+n_1\nu_1+n_2\nu_2) =0$ with integer numbers $n_0$, $n_1$, and $n_2$. The general resonance condition includes multi-photon processes, which indicates the presence of subharmonic interactions, see  \cite{xia2023fast}.  This condition for the case of MPD is simplified to the condition with $\nu_0=\nu_2=0$.

If there is a dominant slow-oscillating component in $G(t)$, this term primarily governs the dynamics of the corresponding transition. Consequently, one can apply the Rotating Wave Approximation (RWA) by disregarding all other fast-oscillating terms.

For MPD  under the resonant condition the static coupling strength between $|100\rangle$ and $|010\rangle$ can be determined as follows:
\begin{equation} \label{eq:g01 10 compact single drive}
\Bar{g}_{100,010}=i\alpha\Omega_1\left[J_2\left(\frac{\beta\Omega_1}{\nu_1}\right)+J_0\left(\frac{\beta\Omega_1}{\nu_1}\right)\right]
\end{equation}
where we define $\alpha\equiv C^1_{100,010}/2$, $\beta \equiv N^1_{100}-N^1_{010}$.

The coupling strength between $|110\rangle$ and $|020\rangle$ can be calculated similarly when the drive frequency $\nu_1$ is resonant with $\Delta_{110,020}=\Tilde{\omega}_{110}-\Tilde{\omega}_{020}$, see Appendix (\ref{appen:G}). This $g$ coupling is imaginary because of the rotating frame we chose. In this frame, the interaction resembles a Y-type rather than an X-type interaction.

The formula Eq.~(\ref{eq:g01 10 compact single drive}) is valid across a wide range, from the dispersive regime to the strong coupling/drive regime, as no perturbative expansion is employed in its derivation. The only approximation utilized is the RWA. The parameters $\alpha$ and $\beta$ are determined by the underlying undriven Hamiltonian and are obtained through exact numerical diagonalization of this Hamiltonian. As a result, our findings are exact with respect to the bare frequencies and bare couplings.

The first term, $\alpha \Omega_1$, represents the dominant linear dependence of $g_{010,100}$ on $\Omega_1$. The constant $\alpha$ is proportional to the coupling coefficient $C^1_{100,010}$. This result naturally arises from the Hamiltonian (\ref{eq:total H diag frame}), where the term associated with $C^1_{100,010}$ is the only static coupling between $|100\rangle$ and $|010\rangle$. Consequently, the leading linear coupling is given by $\alpha \Omega_1$.

The second term, $J_2\left(\beta \Omega_1/\nu_1\right) + J_0\left(\beta \Omega_1/\nu_1\right)$, introduces nonlinearity to $g_{010,100}$. This nonlinearity arises from the modulation of the frequencies of the dressed states $|100\rangle$ and $|010\rangle$ induced by the parametric drives. The argument of the Bessel functions, $\beta \Omega_1/\nu_1$, consists of three components:
\begin{enumerate}
    \item $\beta$ reflects how differently the drives are coupled to the states $|100\rangle$ and $|010\rangle$, given by the difference of the diagonal terms $N^1_{100} - N^1_{010}$.
	\item Amplitude $\Omega_1$ (numerator) determines the drive strength. A stronger drive results in a more significant nonlinear effect.
	\item Frequency $\nu_1$ (denominator) indicates how rapidly the drive modulates the frequencies of the states. Faster modulation reduces the nonlinearity.
\end{enumerate}

This nonlinear contribution is typically absent in perturbative treatments but can become significant when the drive amplitude is comparable to the drive frequency.

We calculate the static coupling strengths $g_{010,100}$ and $g_{110,020}$ under their respective resonant drives ($\nu_1 = \Delta_{010,100}$ and $\nu_1 = \Delta_{110,020}$) and compare the results with numerical simulations in Fig.~(\ref{fig:gs single drive}). The nonlinear behavior is well captured by the analytical formulas. The transition $g_{110,020}$ exhibits more pronounced nonlinearity because the transition frequency $\Delta_{110,020}$ is smaller, leading to a larger ratio $\Omega / \nu$.

The dips observed in Fig.~(\ref{fig:gs single drive}c) and Fig.~(\ref{fig:gs single drive}d) correspond to the condition $J_2\left(\beta \Omega_1/\nu_1\right) + J_0\left(\beta \Omega_1/ \nu_1 \right) = 0$, which results in $g_{110,020} = 0$. Solving the equation $J_2\left(\beta \Omega_1/ \nu_1 \right) + J_0\left(\beta \Omega_1/ \nu_1 \right) = 0$ numerically yields the first root $\beta  \Omega_1/ \nu_1 \approx 3.83$, which matches well with the numerical simulation. Other higher-order roots are beyond the accessible range.

\subsubsection{Drive Crosstalk in BPD}
\label{sec.crosstalk}

\begin{figure*}[ht]
\centering
\begin{subfigure}{0.40\linewidth}
\includegraphics[width=1\textwidth]{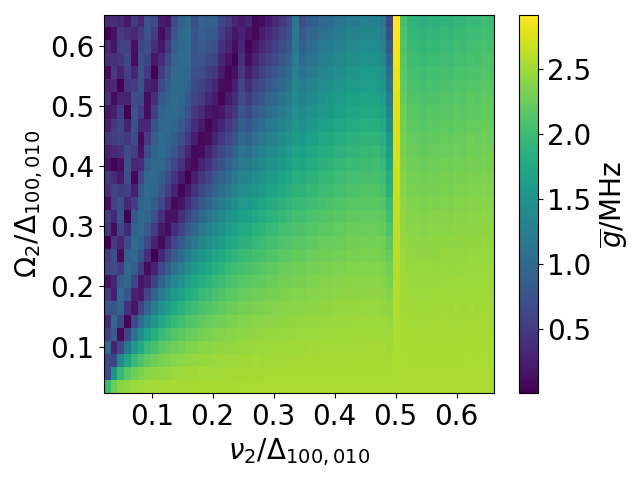}\put(-180,150){\textbf {(a)}}
\end{subfigure}
\begin{subfigure}{0.40\linewidth}
\includegraphics[width=1\textwidth]{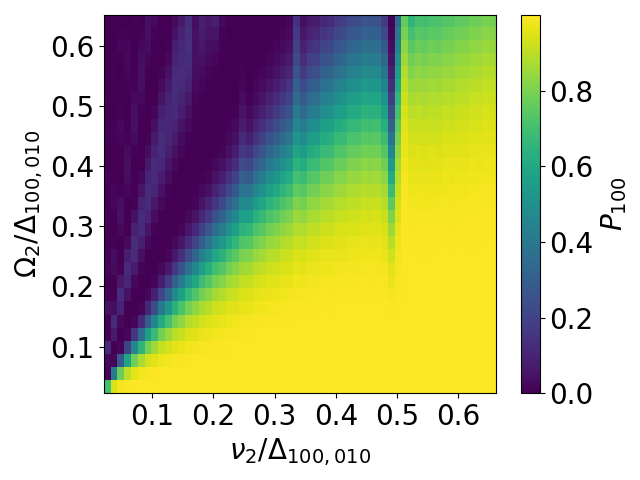}\put(-180,150){\textbf {(b)}}  
\end{subfigure}

\begin{subfigure}{0.40\linewidth}
\includegraphics[width=1\textwidth]{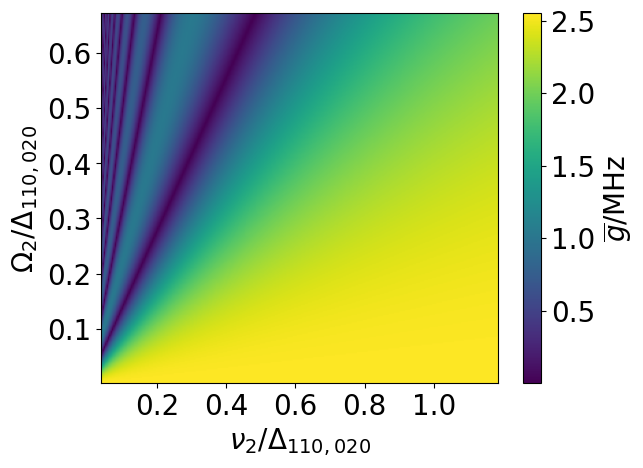}\put(-180,150){\textbf {(c)}}
\end{subfigure}
\begin{subfigure}{0.40\linewidth}
\includegraphics[width=1\textwidth]{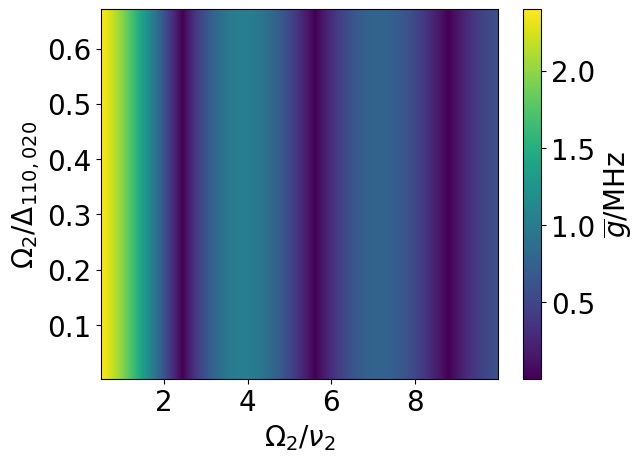}\put(-180,150){\textbf {(d)}}  
\end{subfigure}
\caption{Crosstalk effect of one drive on the other.  Tthe first drive is fixed at $\Omega_1=150$ MHz, $\nu_1=\Delta_{010,100}$. The second drive is far detuned from  $|010\rangle \leftrightarrow |100\rangle$ transition. We sweep drive amplitude $\Omega_2$ and frequency $\nu_2$ to see the crosstalk effects. \ref{fig:g P crosstalk}(a) Numerical simulation of static $\bar{g}_{100,010}$. The static coupling $\bar{g}_{100,010}$ is modulated by the second parametric drive. The dark stripes correspond to zeros of $J_0$ at $\Omega_2/\nu_2=2.40,5.52,8.65,\cdots$.  The anomalous resonance around $\nu_2=225$ MHz is not captured by RWA formula, because at this spot $2\nu_2=\Delta_{010,100}$ and 2-photon transitions need to be included to account for this resonance. \ref{fig:g P crosstalk}(b) Numerical simulation of $P_{100}$ is consistent with  analytical result except for 2-photon transition. \ref{fig:g P crosstalk}(c) Analytical calculation of static $\bar{g}_{010,100}$ using Eq.~(\ref{eq: g01 10 compact cross}). The analytical plot confirms numerical simulation in (a).  \ref{fig:g P crosstalk}(d) Analytical calculation of static $\bar{g}_{010,100}$. The x-axis is rescaled to illustrate the dependence of crosstalk effect on $\Omega_2/\nu_2$. }
\label{fig:g P crosstalk}
\end{figure*}

For the case of BPD with two parametric drives, the static coupling strength $g$ between two states can be evaluated similarly. For instance, the static coupling between $|010\rangle$ and $|100\rangle$ under one resonance drive $\Omega_1 \sin(\nu_1 t)a_1^\dag a_1$ and one off-resonant drive $\Omega_2 \sin(\nu_2 t)a_2^\dag a_2$ is given by:
\begin{equation} \label{eq: g01 10 compact cross}
g_1=i\alpha\Omega_1\left[J_2\left(\frac{\beta\Omega_1}{\nu_1}\right)+J_0\left(\frac{\beta \Omega_1}{\nu_1}\right)\right] J_0\left(\frac{\gamma\Omega_2}{\nu_2}\right)
\end{equation}
where
\begin{equation}\label{eq: alpha}
\alpha=\frac{C^1_{100,010}}{2}, \beta=N^1_{100}-N^1_{010}, \gamma=N^2_{100}-N^2_{010}
\end{equation}

Details of derivation and even more general case of three parametric derives can be found in Appendix (\ref{appen:G}).

The first part of $g_1$, $i\alpha\Omega_1(J_2(\beta \Omega_1/\nu_1) + J_0(\beta\Omega_1/\nu_1))$, is identical to the static $\Bar{g}_{100,010}$ derived under a single resonant drive, as shown in Eq.~(\ref{eq:g01 10 compact single drive}). The crosstalk effect of the off-resonant drive is introduced through the last term, $J_0(\gamma \Omega_2/\nu_2)$. This term takes a similar form to the nonlinear term in the single-drive case. The crosstalk arises from the frequency modulation of the states $|100\rangle$ and $|010\rangle$ induced by the second drive.

The crosstalk factor has two key features:
\begin{enumerate}
    \item Nonlinearity: The crosstalk appears through the nonlinear function $J_0$.
    \item Dependence on the ratio $\Omega_2/\nu_2$: The crosstalk depends solely on this ratio.
\end{enumerate}
	
For small values of $\Omega_2/\nu_2$, $J_0$ is a monotonically decreasing function. As $\Omega_2/\nu_2$ increases, $J_0$ begins to oscillate around zero, resulting in multiple zeros in $\Bar{g}_{100,010}$ and $P_{100}$. This behavior is illustrated in Fig.~(\ref{fig:g P crosstalk}). We numerically compute the static coupling strength $\bar{g}_1$ and the population transfer and compare with the analytical results of Eq.~(\ref{eq: g01 10 compact cross}). It can be clearly seen that the crosstalk is mainly coming through the factor $J_0(\gamma\Omega_2/\nu_2)$.

The crosstalk can be understood as an off-resonant baseband modulation on top of a resonant sideband transition.  The resulting coupling strength is the product of the two parts. Eq.~(\ref{eq: g01 10 compact cross}) can be generalized to N drives, where only one drive is resonant and all other N-1 drives are off-resonant:

\begin{equation} \label{eq.g multi drive}
\Bar{g}=i\alpha\Omega_1(J_2(\beta\frac{\Omega_1}{\nu_1})+J_0(\beta\frac{\Omega_1}{\nu_1}))\Pi_{k=2}^N J_0(\gamma_k\frac{\Omega_k}{\nu_k})
\end{equation}

The formula Eq.~(\ref{eq.g multi drive}) can be derived following the same line as Appendix (\ref{appen:G}). Here we sketch how it can be done. In the rotating frame, the coupling strength takes the form:$g \propto \Sigma_j\alpha_j \Omega_j \sin(\nu_j t)\exp\{i[\nu_1t+\Sigma_k \gamma_k\frac{\Omega_k}{\nu_k}\cos(\nu_k t)]\}$, where $\nu_1$ is the detuning of the targeted transition. One can then expand the exponents using the Jacobi–Anger identity and perform RWA. Because only drive 1 is resonant with the targeted transition and all other drives are off-resonant, the only terms left in the sum belong to the following two cases: 

1. $j=1$ so $\sin(\nu_1 t)$ cancels the phase factor $\exp(i\nu_1t)$

2. $j\neq 1$ but it cancels expansion terms from $\exp(i\gamma_j\frac{\Omega_j}{\nu_j}\cos(\nu_jt))$ and the phase factor $\exp(i\nu_1t)$ is canceled by one expansion term from $\exp(i\gamma_1\frac{\Omega_1}{\nu_1}\cos(\nu_1t))$.

Assuming $J_0(...)>>J_1(...),J_2(...)$, it is easy to find that only the first case gives dominant contribution. This thus gives the result Eq.~(\ref{eq.g multi drive}), where we have renamed $\beta \equiv \gamma_1$. This expression also applies to multi-qubit systems. One can use it to suppress a target transition by requiring $\Pi_{k=2}^N J_0(\gamma_k\frac{\Omega_k}{\nu_k})=0$.

\section{Minimizing Leakage}
\label{sec.Opt}

Driving the CPHASE transition while the iSWAP  drive is present needs more delicate treatment because we require the state always to return to $|11\rangle$ at the end of the gate with a fixed gate time $t_g$. When one of the drives is far-detuned, $|\Delta_{110,020}-\nu_1|\gg\Delta_{110,020}$, and the other is near-resonant $|\Delta_{110,020}-\nu_2|\ll\Delta_{110,020}$. The first order coupling strength between $|110\rangle$ and $|020\rangle$ is:

\begin{equation} \label{eq: g11 02 compact}
\begin{split}
g_2=i&\alpha\Omega_2J_0\left(\frac{\beta\Omega_1}{\nu_1}\right)  \left[J_2\left(\frac{\gamma\Omega_2}{\nu_2}\right)+J_0\left(\frac{\gamma\Omega_2}{\nu_2}\right)\right]\\
    & \times \exp \left({i(\Tilde{\Delta}_{110,020}-\nu_2 )t} \right)
\end{split}
\end{equation}
with the following definitions:
\begin{equation}\label{eq: alpha 11 20}
\alpha=\frac{C^2_{110,020}}{2}, \beta=N^1_{110}-N^1_{020}, \gamma=N^2_{110}-N^2_{020}
\end{equation}

Note that Eq.~(\ref{eq: g11 02 compact}) has similar structure as Eq.~(\ref{eq: alpha}). The only difference is that because here we assume the drive frequency $\nu_2$ is slightly detuned from $\Tilde{\Delta}_{110,020}$, there is an extra slow-varying phase factor $\exp \left({i(\Tilde{\Delta}_{110,020}-\nu_2 )t} \right)$, details can be found in Appendix (\ref{appen:G})

Within the space of acting CPHASE, which are $\{|110\rangle, |020\rangle\}$ subspace, we can define the following Pauli operators:

\begin{equation}\label{eq: Pauli 11 20}
    \begin{split}
        &\hat{\Sigma}^+=|020\rangle\langle110|, \quad 
        \hat{\Sigma}^-=|110\rangle\langle020|\\
        &\hat{\Sigma}^z=|110\rangle\langle110|-|110\rangle\langle020|\\
        &\hat{\Sigma}^x=|020\rangle\langle110|+|110\rangle\langle020|
    \end{split}
\end{equation}

The effective Hamiltonian in this subspace can then be written as:

\begin{equation}\label{eq: eff H 11 20}
    \begin{split}
        H=ig_2\hat{\Sigma}^+e^{i\Delta t}+H.c.
    \end{split}
\end{equation}
with the definition of $g$ as follows:
\begin{equation}\label{eq: eff H 11 20 notions}
    \begin{split}
        \Delta&=\Delta_{110,020}-\nu_2\\
        g_2&=\alpha\Omega_2J_0\left(\frac{\beta\Omega_1}{\nu_1}\right)\left(J_2\left(\frac{\gamma \Omega_2}{\nu_2}\right)+J_0\left(\frac{\gamma\Omega_2}{\nu_2}\right)\right)
    \end{split}
\end{equation}

We can now make another rotating frame transformation $U_r=\exp[{-i(\Delta t+\pi/2)\hat{\Sigma}^z}/2]$ to rotate away the time-dependent factor $ie^{i\Delta t}$. The transformed Hamiltonian is given by :
\begin{equation}\label{eq: eff H 11 20 final}
    \begin{split}
        \Tilde{H}&=U_r^\dag H U_r -iU_r^\dag \frac{\partial U_r}{\partial_t}\\
        &=\frac{\Delta}{2}\hat{\Sigma}^z+g_2\hat{\Sigma}^x
    \end{split}
\end{equation}

The time evolution can now easily be calculated:

\begin{equation} \label{eq: U fsim 11 20}
\begin{split}
U(t)&=e^{-i\Tilde{H}t}\\
&=\cos(\Omega t) \hat{I}-i\sin(\Omega t) (n_z\hat{\Sigma}^z+n_x\hat{\Sigma}^x)
\end{split}
\end{equation}

where $\Omega= \sqrt{(\frac{\Delta}{2})^2+g_2^2}, n_z={\Delta}/{\sqrt{\Delta^2+4g_2^2}} $ and $n_x={2g_2}/{\sqrt{\Delta^2+4g_2^2}}$. Note that a part of conditional phase due to ZZ interaction is now absorbed in the rotating frame transformation $U_r$ \cite{Ku2020}.

To maintain a constant gate time $t_g$, we require:

\begin{equation} \label{eq: fsim opt amp condition}
\begin{split}
\Omega  t_g&= \pi \\
\Rightarrow g&=\pm \sqrt{(\frac{\pi}{t_g})^2-(\frac{\Delta}{2})^2}
\end{split}
\end{equation}

Now combining Eq.~(\ref{eq: eff H 11 20 notions}) and Eq.~(\ref{eq: fsim opt amp condition}), we can numerically solve for the optimal amplitude of the second drive $\Omega_2$ for a given set of $(t_g, \nu_1, \Omega_1, \nu_2)$:

\begin{widetext}
\begin{equation} \label{eq:fsim opt amp final eq}
\begin{split}
\alpha\Omega_2J_0\left(\beta\frac{\Omega_1}{\nu_1}\right)\left(J_2\left(\gamma\frac{\Omega_2}{\nu_2}\right)+J_0\left(\gamma\frac{\Omega_2}{\nu_2}\right)\right)=\pm \sqrt{\left(\frac{\pi}{t_g}\right)^2-\left(\frac{\Delta_{110,020}-\nu_2}{2}\right)^2}
\end{split}
\end{equation}    
\end{widetext}

\begin{figure*}[t]
\centering
\begin{subfigure}{0.32\linewidth}
    \includegraphics[width=\linewidth]{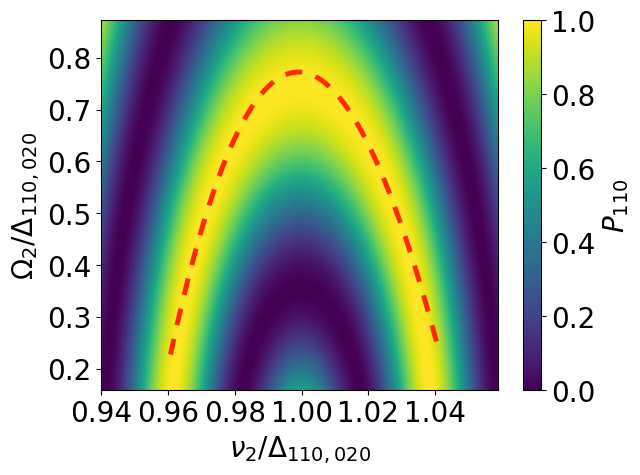}
    \put(-155,120){\textbf{(a)}}
\end{subfigure}
\hfill
\begin{subfigure}{0.32\linewidth}
    \includegraphics[width=\linewidth]{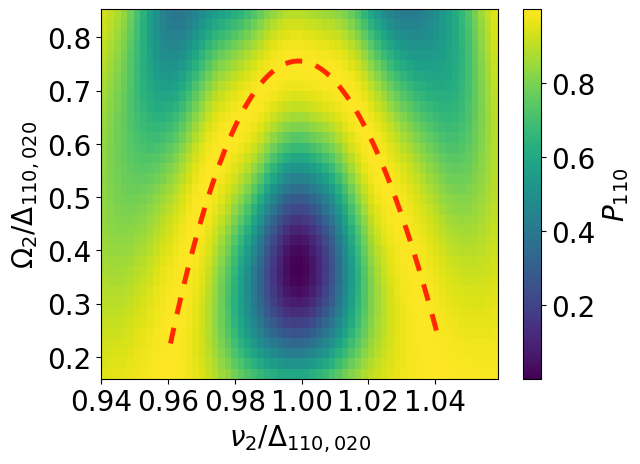}\put(-155,120){\textbf{(b)}}
\end{subfigure}
\hfill
\begin{subfigure}{0.32\linewidth}
    \includegraphics[width=\linewidth]{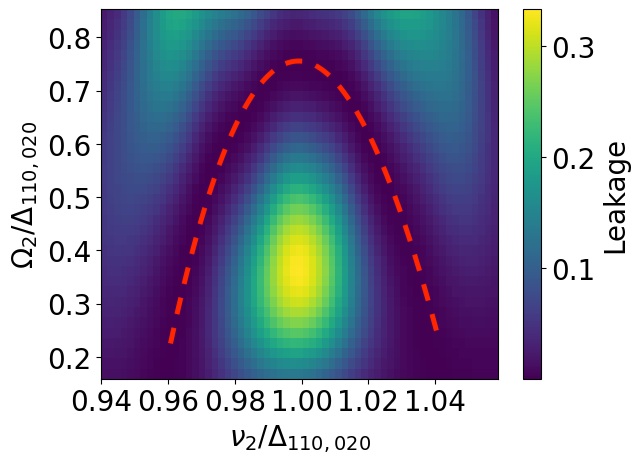}\put(-155,120){\textbf{(c)}}
\end{subfigure}
\caption{Identify the optimal amplitude trajectory. The optimal trajectory is marked by the red dashed line. (a) Analytical plot of P11 vs $\Omega_2$ and $\nu_2$ while the first drive is resonant with $|01\rangle \leftrightarrow |10\rangle$ transition: $\Omega_1=150$ MHz, $\nu_1=\Delta_{01,10}$. The optimal drive amplitude $\Omega_2$ is defined as such that P11 is maximum. (b) Numerical plot of P11 vs $\Omega_2$ and $\nu_2$. It is consistent with the analytical plot. (c) Numerical plot of leakage vs $\Omega_2$ and $\nu_2$. We calculate the average leakage out of the computational subspace when the initial states are $|010\rangle$, $|100\rangle$ and $|110\rangle$. The optimal drive amplitude $\Omega_2$ can also be defined as such to minimize the leakage.}
\label{fig:trajectory num and ana}
\end{figure*}

Eq.~(\ref{eq:fsim opt amp final eq}) is the main equation that we need for determining the optimal amplitude of the second drive. Note that once $\alpha, \beta$ and $\gamma$ are fixed in Eq.~(\ref{eq: eff H 11 20 notions}) by a given undriven Hamiltionian, the sign of $g_{1,2}$ normally won't change within a reasonable range of $\frac{\nu_{1,2}}{\Omega_{1,2}}$ unless we flip the signs of $\Omega_{1,2}$ or $\nu_{1,2}$. So we will only see a single branch in the numerical solution of $g_{1,2}$. The optimal amplitude shows up as a low-leakage trajectory around a high-leakage center in the 2D scan of $\nu_2$ and $\Omega_2$. Following the trajectory we have the lowest leakage ( complete cycle of $|110\rangle \rightarrow |020\rangle \rightarrow |110\rangle$) and also tunable CPHASE, see Fig.~\ref{fig:trajectory num and ana}.

\begin{figure}
\includegraphics[width=0.4\textwidth]{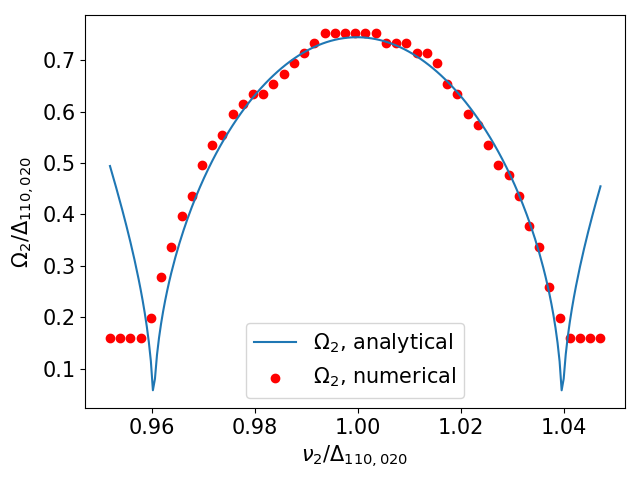}\put(-200,150){\textbf {(a)}}
\hfill
\includegraphics[width=0.4\textwidth]{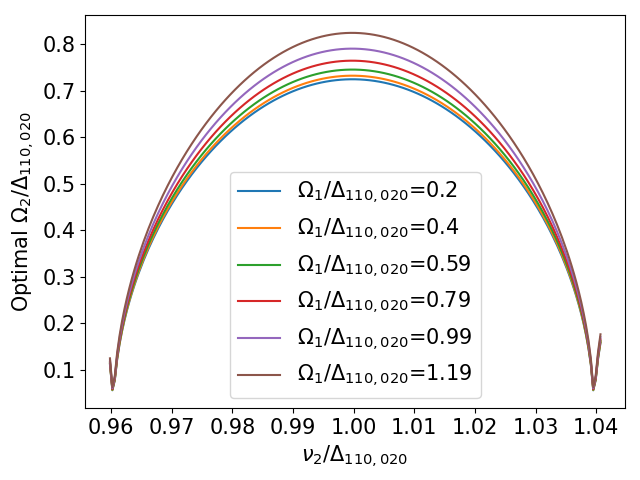}\put(-200,150){\textbf {(b)}}
 
\caption{Optimal $\Omega_2$ vs $\nu_2$. \ref{fig:opt amp arch ana and num}(a) numerically determined optimal drive amplitude $\Omega_2$ by minimizing leakage and analytical optimal $\Omega_2$. The discrepancy on the edges is because $|\nu_2-\Delta_{110,020}|>\frac{2\pi}{t_g}$. No valid solution of effective coupling $g_{110,020}$ can be found to satisfy the condition that one and only one cycle of $|110\rangle \rightarrow |020\rangle \rightarrow |110\rangle$ is completed in the given gate time $t_g$. Therefore, only the $\nu_2$ and $\Omega_2$ in the middle area are relevant. \ref{fig:opt amp arch ana and num}(b) Analytically calculated optimal $\Omega_2$ vs $\nu_2$ with various $\Omega_1$. }
\label{fig:opt amp arch ana and num}
\end{figure}

The analytical formula Eq.~(\ref{eq:fsim opt amp final eq}) can accurately predict the optimal drive amplitude $\Omega_2$, Fig.~(\ref{fig:opt amp arch ana and num}). The curve for optimal $\Omega_2$ has the shape of a semi-ellipse. In fact, up to first order approximation, Eq.~(\ref{eq:fsim opt amp final eq}) becomes $(\alpha J_0(\beta\frac{\Omega_1}{\nu_1})\Omega_2)^2+(\frac{\Tilde{\Delta}_{110,020}-\nu_2}{2})^2=(\frac{\pi}{t_g})^2$, which defines an elliptic curve. We also notice that for a given $\nu_2$, a larger $\Omega_1$ requires a larger $\Omega_2$, see Fig. (\ref{fig:opt amp arch ana and num} b). This is because for a given $\nu_2$ the target $g$ is fixed, see Eq.(\ref{eq: fsim opt amp condition}). However, due to the crosstalk effect of the first drive, one needs to increase $\Omega_2$ to compensate for the modulation factor $J_0(\beta\frac{\Omega_1}{\nu_1})$. This is consistent with the result in Fig. (\ref{fig:g P crosstalk}).

\section{Concurrency in the cfSim gate}\label{sec:cfsim}

In this section, we will show how to achieve full range control of the iSWAP angle and conditional phase by using BPD in one run. The simultaneous drive scheme in principle also applies to other drive combinations. In Appendix (\ref{Appen:static + p fsim}), we show simulations of simultaneous resonant iSWAP and parametric CPHASE, where the two qubits are tuned into resonance and a parametric drive is also present for the 11-02/20 transition. In the main text, we focus on the BPD scheme. Combining the results above, we can now write down the equations for the iSWAP  angle $\theta$ and conditional phase $\varphi$ including the contribution from ZZ interaction and the constraint condition for $\Omega_2$ for a given set of $(t_g, \Omega_1, \nu_2$) with $\nu_1=\Delta_{100,010}$:
\begin{equation}\label{eq: dominant eqs: iswap  cphase opt amp}
    \begin{split}
    &\sin(\theta)=\frac{1-\cos(2g_1t_g)}{2}\\
    &\varphi =\pi(1-\frac{\Delta_{110,020}-\nu_2+2\xi _{zz}}{2\pi}t_g)\\
    &g_2=\pm \sqrt{(\frac{\pi}{t_g})^2-(\frac{\Delta_{110,020}-\nu_2}{2})^2}
    \end{split}
\end{equation}

with

\begin{equation}\label{eq: dominant eqs g}
    \begin{split}
    &g_1=\alpha_1\Omega_1\left(J_2\left(\beta_1\frac{\Omega_1}{\nu_1}\right)+J_0\left(\beta_1\frac{\Omega_1}{\nu_1}\right)\right)J_0\left(\gamma_1\frac{\Omega_2}{\nu_2}\right)\\
    &g_2=\alpha_2\Omega_2J_0\left(\beta_2\frac{\Omega_1}{\nu_1}\right)\left(J_2\left(\gamma_2\frac{\Omega_2}{\nu_2}\right)+J_0\left(\gamma_2\frac{\Omega_2}{\nu_2}\right)\right)
    \end{split}
\end{equation}

where $\alpha_{1,2},\beta_{1,2},\gamma_{1,2}$ are constants given by the underlying undriven Hamiltonian, see Eq.~(\ref{eq: alpha}) and Eq.~(\ref{eq: alpha 11 20}). $\xi_{zz}$ is the ZZ interaction strength given by:

\begin{equation}\label{eq: ZZ}
    \xi_{zz}=E_{110}+E_{000}-E_{100}-E_{010}
\end{equation}

In the numerical simulation, we use the following parameters from Ref.\cite{strand2013first}:

\

\begin{tabularx}{0.4\textwidth} { 
  | >{\centering\arraybackslash}X 
  | >{\centering\arraybackslash}X 
  | >{\centering\arraybackslash}X | }
 \hline
 $\omega_1/2\pi$ & $\omega_2/2\pi$ & $\omega_c/2\pi$ \\
 \hline
 7.15 GHz  & 7.6 GHz  & 8.5 GHz  \\
\hline
\end{tabularx}

\

\begin{tabularx}{0.4\textwidth} { 
  | >{\centering\arraybackslash}X 
  | >{\centering\arraybackslash}X 
  | >{\centering\arraybackslash}X | }
 \hline
 $\delta_1/2\pi$ & $\delta_2/2\pi$ & $\delta_c/2\pi$ \\
 \hline
 -200 MHz  & -200 MHz  & 0  \\
\hline
\end{tabularx}

\

\begin{tabularx}{0.4\textwidth} { 
  | >{\centering\arraybackslash}X 
  | >{\centering\arraybackslash}X 
  | >{\centering\arraybackslash}X | }
 \hline
 $g_{1c}/2\pi$ & $g_{2c}/2\pi$ & $g_{12}/2\pi$ \\
 \hline
 120 MHz  & 120 MHz  & 0  \\
\hline
\end{tabularx}

\

\begin{tabularx}{0.4\textwidth} { 
  | >{\centering\arraybackslash}X 
  | >{\centering\arraybackslash}X 
  | >{\centering\arraybackslash}X | }
 \hline
 $\phi_1^d$ & $\phi_2^d$ & $t_g$ \\
 \hline
 0  & 0  & 100 ns  \\
\hline
\end{tabularx}

\hfill

\begin{figure*}[!ht]
\centering
\begin{subfigure}{0.4\linewidth}
\includegraphics[width=1\textwidth]{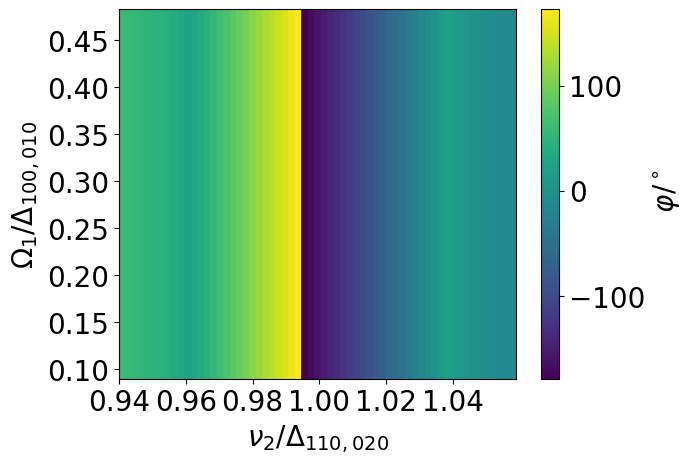}\put(-185,135){\textbf {(a)}}
\end{subfigure}
\begin{subfigure}{0.4\linewidth}
\includegraphics[width=1\textwidth]{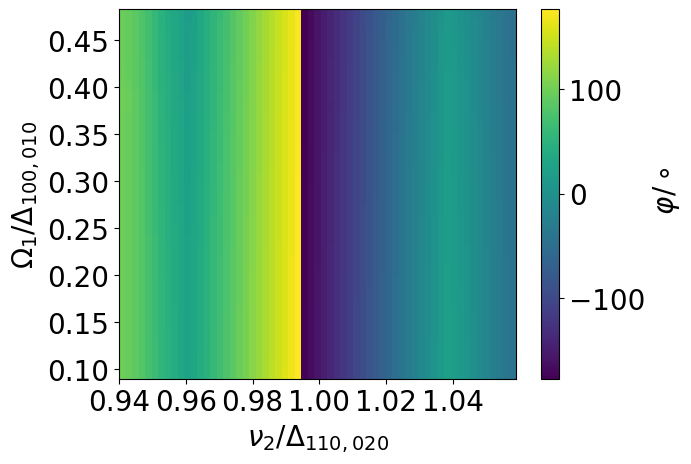}\put(-185,135){\textbf {(b)}}
\end{subfigure}\\
\begin{subfigure}{0.4\linewidth}
\includegraphics[width=1\textwidth]{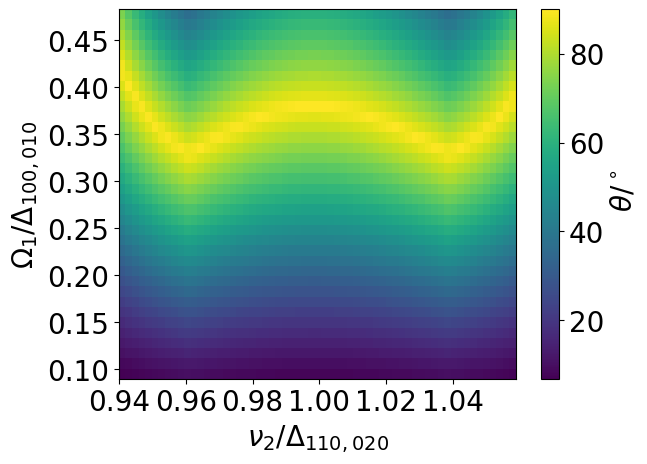}\put(-185,145){\textbf {(c)}}
\end{subfigure}
\begin{subfigure}{0.4\linewidth}
\includegraphics[width=1\textwidth]{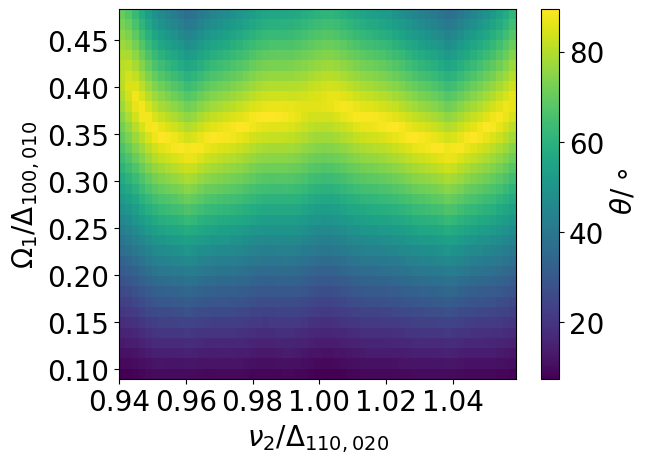}\put(-185,145){\textbf {(d)}}
\end{subfigure}

\caption{Continuous fSim gate. Both $\theta$ and $\varphi$ are continuously tunable. \ref{fig:fSim num ana}(a) Analytically calculated conditional phase vs $\Omega_1$ and $\nu_2$. The phase covers the full $2\pi$ range from $-\pi$ to $\pi$. The phase is dominantly determined by $\nu_2$. The vertical cut in the middle is because the range is chosen to be $[-\pi,\pi)$ in order to manifest the zero conditional phase trajectory. Along the vertical line $\varphi = \pi(-\pi)$. \ref{fig:fSim num ana}(b) Numerically calculated conditional phase vs $\Omega_1$ and $\nu_2$. It agrees with the analytical plot well. \ref{fig:fSim num ana}(c) Analytically calculated iSWAP  angle $\theta$ vs $\Omega_1$ and $\nu_2$.  $\theta$ is tunable from 0 to $90^\circ$. $\theta$ is mainly controlled by $\Omega_1$ with some modulation from $\nu_2$. A larger $\Omega_1$ is needed to achieve maximum $\theta$ when $\nu_2$ is resonant with $\Delta_{110,020}$. This is because when $\nu_2$ gets closer to $\Delta_{110,020}$, the corresponding optimal $\Omega_2$ becomes larger, which in turn makes $\frac{\Omega_2}{\nu_2}$ larger. The effective coupling $g_{010,100}$ is thus reduced, see Eq.~(\ref{eq: g01 10 compact cross}) and Fig.~(\ref{fig:g P crosstalk}). \ref{fig:fSim num ana}(d) Numerically calculated iSWAP  angle $\theta$ vs $\Omega_1$ and $\nu_2$, consistent with analytical plot.}
\label{fig:fSim num ana}
\end{figure*}

We first calculate the optimal drive amplitude $\Omega_2$ according to Eq.~(\ref{eq:fsim opt amp final eq}). We then use the analytically calculated $\Omega_2$ in the time-dependent Schr\"{o}dinger equation and numerically solve the equation. Within the valid area of Eq.~(\ref{eq:fsim opt amp final eq}), the leakage is suppressed below 1\%.
Note that in Eq.~(\ref{eq: ZZ}) we have assumed that the coupler $|R\rangle$ is in the ground state. This is valid because the coupler is far detuned from both qubits in GHz order. During the gate, no drive is resonant with any coupler excitation. So the coupler won't be occupied during the gate 
execution.
In Fig.~(\ref{fig:fSim num ana}), we plot the iSWAP  angle $\theta$ and conditional phase $\varphi$ from numerical simulation and compare it to analytical results from Eq.~(\ref{eq: dominant eqs: iswap  cphase opt amp}). $\theta$ is tunable from 0 to $90^\circ$ and $\varphi$ is tunable from $-180^\circ$ to $180^\circ$. Both can be faithfully predicted by the analytical formulas. To quantify the effect of leakage, we calculate the gate fidelity using the formula\cite{pedersen2007fidelity}:

\begin{equation}
    F(U,U_0)=\frac{Tr(MM^\dag)+|Tr(M)|^2}{d(d+1)}
\end{equation}

where $M=PU_0UP$. $U$ is the unitary operator of the gate process. $U_0$ is the ideal gate operator. $P$ is the projection operator onto the computational subspace spanned by $\{|000\rangle, 
|100\rangle, |010\rangle, |110\rangle\}$ and $d$ is the dimension of $U$. 

\begin{figure*}[!ht]
\includegraphics[width=0.35\textwidth]{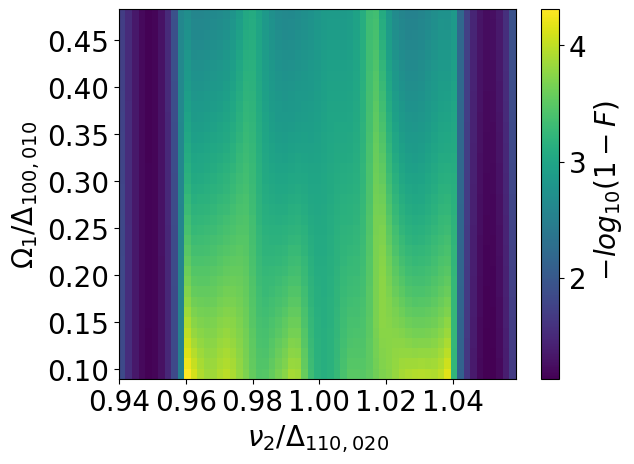}\put(-165,135){\textbf {(a)}}
\includegraphics[width=0.35\textwidth]{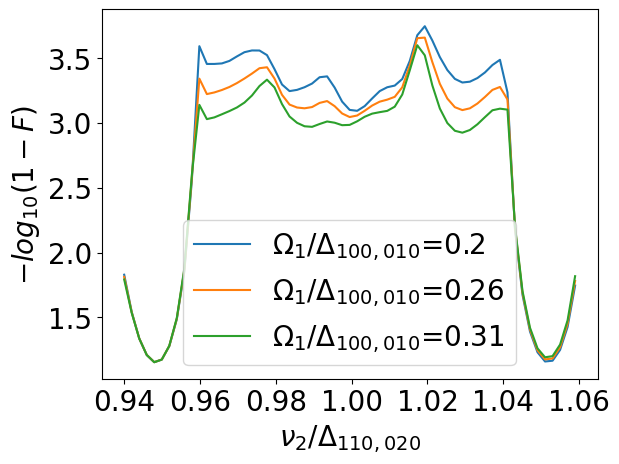}\put(-165,135){\textbf {(b)}}\\
\includegraphics[width=0.35\textwidth]{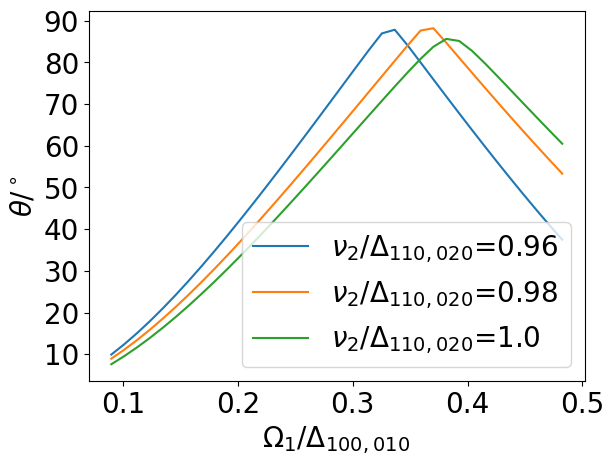}\put(-175,135){\textbf {(c)}}
\includegraphics[width=0.35\textwidth]{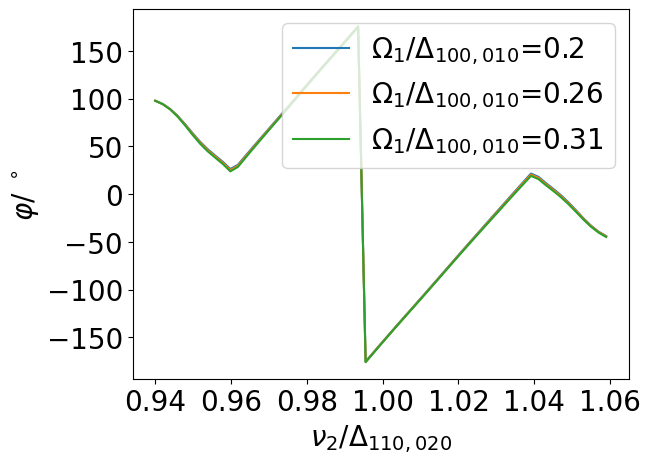}\put(-165,130){\textbf {(d)}}
        
\caption{\ref{fig:fSim num fide log and linecut}(a) Fidelity of the fSim gate. The dark area is out of the fSim gate range. Within the fSim gate range, the fidelity is around or above 99.9\%. \ref{fig:fSim num fide log and linecut}(b) Fidelity along the $\nu_2$ axis while $\Omega_1$ is fixed. Three line cuts are shown. The fidelity domain is between 99.9\% to 99.99\% , subject to qubit decoherence.  \ref{fig:fSim num fide log and linecut}(c) iSWAP  angle $\theta$ vs $\Omega_1$ while $\nu_2$ is fixed. \ref{fig:fSim num fide log and linecut}(d) Conditional phase $\varphi$ vs $\nu_2$ while $\Omega_1$ is fixed. The crosstalk effect on the conditional phase is small. }
\label{fig:fSim num fide log and linecut}
\end{figure*}

The fidelity from the numerical simulation is shown in Fig.~(\ref{fig:fSim num fide log and linecut}a). In the fSim gate range, the fidelity is consistently above 99.5\%. In most area it's above 99.9\%. The cross-talk effect can be clearly seen in the line cuts Fig.~(\ref{fig:fSim num fide log and linecut}c). The gate fidelity is lower when $\Omega_1$ is larger. This is most likely because of the beyond-RWA effect, which is left for future study. This fidelity is at the same level as that given by common two-qubit gate schemes such as iSWAP or CZ. Nevertheless, our flexible concurrent fSim gates can reduce circuit depth compared to the one-native-2Q-gate approach. This can improve the overall fidelity of a quantum circuit.

The gate fidelity can be further improved by including pulse-shaping. Here we choose to use flat-top Gaussian to demonstrate the effect of pulse-shaping. One crucial step in our numerical simulation is to find the optimal drive amplitude $\Omega_2$, see Eq.~(\ref{eq: dominant eqs: iswap  cphase opt amp}) and Eq.~(\ref{eq: dominant eqs g}). This was calculated by using RWA when the drives are rectangular pulses. In the case of pulse-shaping, however, the RWA result is inaccurate because of the slow-changing rise-and-fall edges, especially when one wants to achieve high gate fidelity. In order to overcome this problem, we choose to make a two-level approximation and numerically optimize the drive amplitude in the effective two-level system, which is spanned by $|110\rangle$ and $|020\rangle$ in our case. One can also include more relevant levels accordingly. During the numerical optimization, only the pulse amplitude is varied. Other parameters, such as pulse width and rise-and-fall time, are fixed, which can also affect the gate fidelity. The pulse envelope to be optimized is therefore expressed as:
\begin{equation}\label{eq: op factor}
    e(t)=\Gamma f(t)
\end{equation}

$e(t)$ is the final pulse envelope and $f(t)$ is the standard flat top Gaussian envelope.
$\Gamma$ is the parameter to be optimized. The flat top Gaussian $f(t)$ is defined as:
\begin{equation}
f(t)=\begin{cases}
        \exp(-\frac{(t-\tau_1)^2}{2\sigma^2}) & \text{,  } 0<t<\tau_1\\
        1 & \text{, } \tau_1<t<\tau_1+\tau_2\\
        \exp(-\frac{(t-\tau_1-\tau_2)^2}{2\sigma^2}) & \text{,  } \tau_1+\tau_2<t<2\tau_1+\tau_2
    \end{cases}
\end{equation}

\begin{figure}[!ht]
\includegraphics[width=0.48\textwidth]{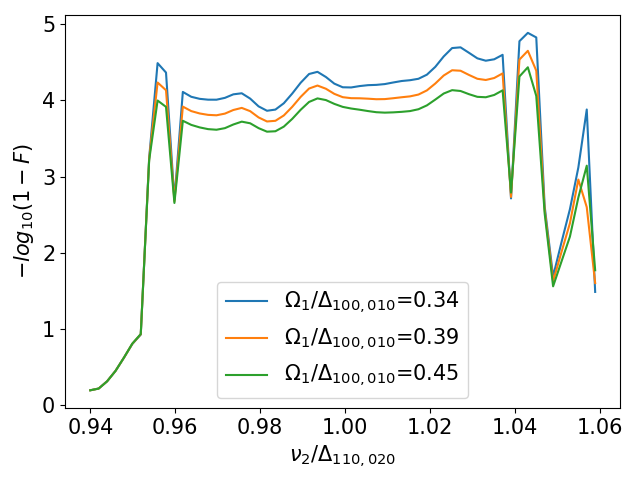}   
\caption{ Three specific line cuts along the $\nu_2$ axis of fSim gate fidelities with flat-top Gaussian
pulse-shaping. The fidelities are improved by almost one order of
magnitude. We use larger drive amplitudes $\Omega_1$ than in Fig. (\ref{fig:fSim num fide log and linecut}b) because it requires stronger drive to achieve the same iSWAP angles after adding pulse-shaping. }
\label{fig:fsim PS fid linecut}
\end{figure}

We find the optimal drive amplitude by requiring minimal leakage to $|020\rangle$, or equivalently maximal $|110\rangle$ occupation. We find that the gate fidelity is improved by approximately one-order-of-magnitude. The iSWAP and CPHASE patterns are roughly unchanged. To illustrate the enhancement to fidelity, we show three specific fidelity trajectories versus $\nu_2$ in Fig.~(\ref{fig:fsim PS fid linecut}).

Nevertheless, we do find more discontinuities/jumps in the plots. This is because of the numerical optimization. During the optimization, especially when the drive frequency $\nu_2$ is approaching the boundaries of the valid area of fSim gates, the resulting optimization factors may not be continuous in contrast to the analytical formulas Eq.~(\ref{eq: dominant eqs: iswap  cphase opt amp}) and Eq.~(\ref{eq: dominant eqs g}) used when there is no pulse-shaping.

We notice that the CPHASE gate can also be driven by multi-photon processes. According to Eq.~(\ref{eq:res exact2}), any transition defined by $\Delta_{110,\eta}-n_0 \nu_0-n_1 \nu_1-z\nu_2 =0$ can be used to implement the conditional phase, where $\eta$ denotes an ancillary level.

\section{$ZZ$-free iSWAP in the cfSim}\label{sec.ZZ free}

As an example of the concurrent fSim gate, we show how to implement a $90^\circ$ iSWAP gate with zero ZZ phase. We first solve for $(\nu_2,\Omega_1,\Omega_2)$ needed to achieve $\theta=90^\circ,\varphi=0$. In fact, once $\theta, \varphi$ are given, one can analytically solve for $g_1,g_2$ and $\nu_2$ using Eq.~(\ref{eq: dominant eqs: iswap  cphase opt amp}).  We first notice that $\theta$ is only dependent on $g_1$. The conditional phase $\varphi$ and $g_2$ are both only dependent on $\nu_2$. We can, therefore, first invert the expressions of  $\theta$ and $\varphi$. After having $\nu_2$ expressed as a function of $\varphi$, we can substitute it into the expression of $g_2$ to get the equation between $g_2$ and $\varphi$ . The resulting equations are:

\begin{equation}\label{eq:solve for given theta phi}
    \begin{split}
        &g_1=\frac{\arccos(1-2\sin\theta)}{2t_g}\\
        &\nu_2=\Delta_{110,020}-2(\frac{\pi-\varphi}{t_g}-\xi_{zz})\\
        &g_2=\pm \sqrt{(\frac{\pi}{t_g})^2-(\frac{\pi-\varphi}{t_g}-\xi_{zz})^2}
    \end{split}
\end{equation}

\begin{figure}[!ht]
\includegraphics[width=0.24\textwidth]{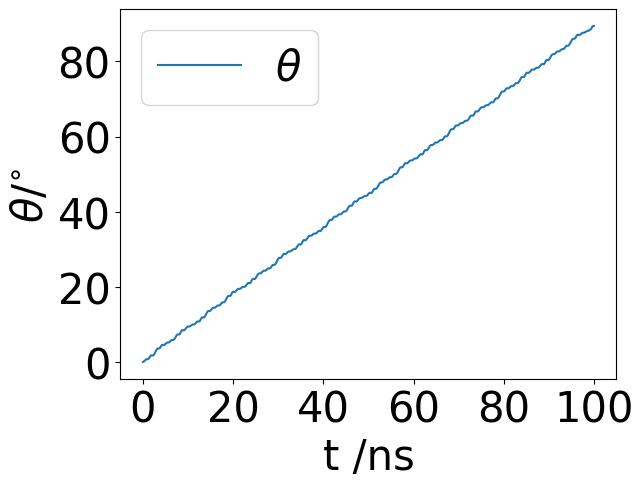}\put(-120,90){\textbf {(a)}}
\includegraphics[width=0.24\textwidth]{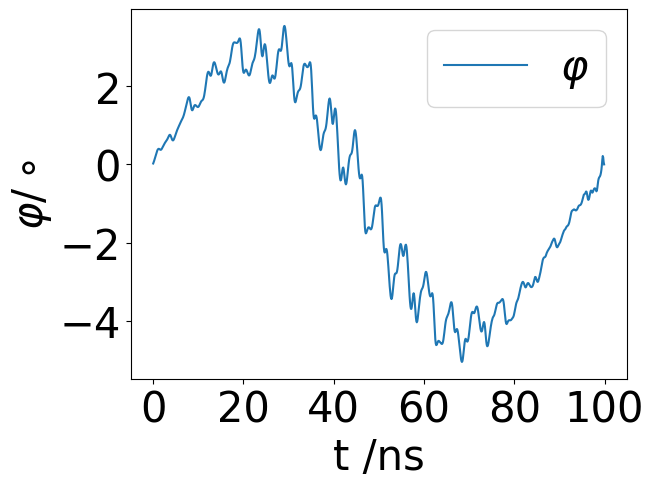}\put(-120,90){\textbf {(b)}}\\

\includegraphics[width=0.24\textwidth]{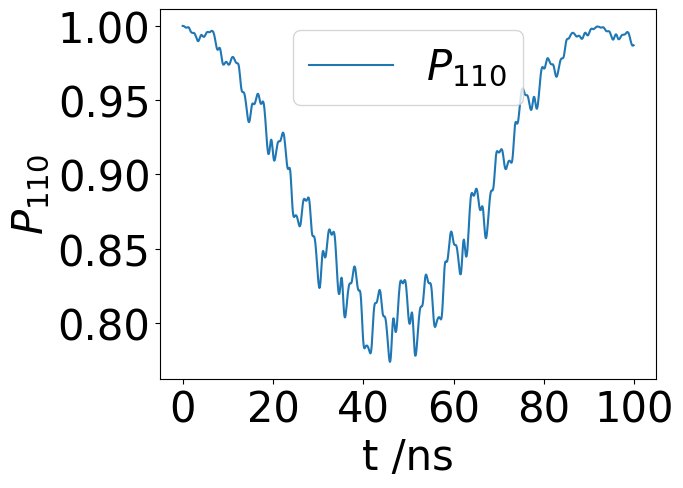}\put(-110,90){\textbf {(c)}}
\includegraphics[width=0.24\textwidth]{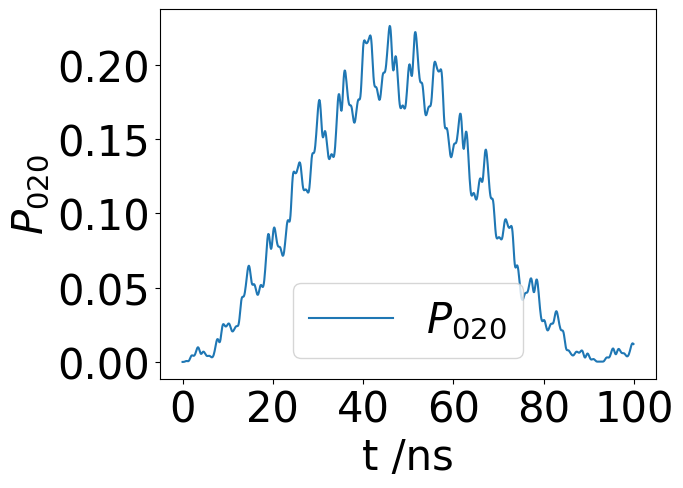}\put(-130,90){\textbf {(d)}}

\caption{An example showing pure iSWAP gate (conditional phase $\varphi=0$). Here we choose full iSWAP  $\theta = 90^\circ$. Fidelity= 99.67$\%$. \ref{fig:zero zz}(a) Real-time evolution of iSWAP  angle $\theta$. \ref{fig:zero zz}(b) Dynamics of conditional phase. It is governed by both the ZZ interaction and geometric phase. The overall phase is zero at the end of the gate. \ref{fig:zero zz}(c) Evolution of $P_{110}$. It corresponds to a closed trajectory in the Bloch sphere. There is a bit of over-rotation which causes leakage. This is because we used rotating wave approximation in the analytical formula. This over-rotation can be corrected by including higher-order oscillating terms. \ref{fig:zero zz}(d) Evolution of $P_{020}$. There is some residual population at the end of the evolution.}
\label{fig:zero zz}
\end{figure}

Therefore, one only needs to numerically solve Eq.~(\ref{eq: dominant eqs g}) for $(\Omega_1,\Omega_2)$. In Fig.~(\ref{fig:zero zz}), we show the real-time evolution of the system. The target angles, $\theta=90^\circ,\varphi=0$ are achieved at the end of the gate. We notice that there is a bit of over-rotation in $P_{110}$ and $P_{020}$. This is because of the beyond-RWA effect. It can be corrected by including fast oscillating terms in Eq.~(\ref{eq8}).

When the qubit is in idle mode, one can use the same method to eliminate ZZ phase. Assuming a far-off-resonant drive $\Omega \sin(\nu t)$. The oscillation period is given by $t_g=\frac{2\pi}{\sqrt{\Delta^2+4g^2}}$, where $\Delta=\Delta_{110,020}-\nu$. The accumulated phase is $\varphi=\pi(1-\frac{\Delta }{\sqrt{\Delta^2+4g^2}})-\xi_{zz}t_g $. Requiring $\varphi=0$ we arrive at:

\begin{equation}\label{eq:zero zz idle}
    g^2-\xi_{zz}\Delta=\frac{\xi_{zz}^2}{4}
\end{equation}

One can numerically solve Eq.~(\ref{eq:zero zz idle}) by combining $g=\alpha\Omega*(J_2(\beta\frac{\Omega}{\nu})+J_0(\beta\frac{\Omega}{\nu}))$.  Our method, in this case, can be applied to cancel static ZZ crosstalk in superconducting quantum processors.

\section{Discussion and outlook}\label{sec.Discussion and outlook}

We have shown that our BPD scheme can efficiently and independently control iSWAP operation and CPHASE operation with high fidelity. Because we focus on the unitary dynamics behind BPD drive, we have left out the crucial topic of optimal gate time considering trade-off between unitary errors and decoherence.

Unitary errors such as leakage usually can be suppressed by prolonging gate time, but the gate time upper bound is set by T1 and T2 because at some point decoherence becomes the dominant error source. Faster gates can reduce decoherence while they suffer from increased unitary errors. In our analysis, we have mainly focused on reducing leakage and neglected most of non-RWA effects. Our numerical simulation of Hamiltonian Eq.~(\ref{eq:setup}) is exact. There is no extra approximation used other than discretizing the time-dependent Hamiltonian into piecewise  constant Hamiltonians with small time steps and truncating the Hilbert space to 100 levels ( 5 levels for each transmon and 4 levels for the coupler). In this way, we partially capture non-RWA and other nonlinear effects. However, as our starting point is the RWA Hamiltonian Eq.~(\ref{eq:setup}), most of the non-linear effects are already approximated away from the very beginning. To fully address all strong drive effects, one needs to use the full circuit Hamiltonian as in Eq.~(\ref{eq:full circuit H}) and Eq.~(\ref{eq:full circuit H in n p}). In fact, in another comprehensive study of strong drives on transmons, it is found that strong driving can ionize transmons and the system enters a regime of chaos\cite{cohen2023reminiscence}. Here we provide some preliminary simulation and study of the non-RWA and other non-linear effects without using the full circuit Hamiltonian model. 

Instead of the RWA Hamiltonian Eq.~(\ref{eq:setup}), we include counter-rotating terms to get a non-RWA Hamiltonian:

\begin{equation}\label{eq:setup non RWA}
    \begin{split}
    &  H=H_0+H_d,\\ 
        & H_0=\sum_{\alpha=1,2,c} \omega_\alpha \adag_\alpha \ah_\alpha + \frac{\delta_j}{2} \adag_\alpha \ah_\alpha (\adag_\alpha \ah_\alpha-1) \\ & \quad \quad \quad + \sum_{j,k=\{1,2\}} g_{jc} (\adag_j+\hat{a}_j)( \adag_c+\hat{a}_c),\\
        & H_d=\sum_{j=1,2} \Omega_j \sin(\nu_j t + \phi^d_j) \adag_j \ah_j.
    \end{split}
\end{equation}

\begin{figure}
\includegraphics[width=0.4\textwidth]{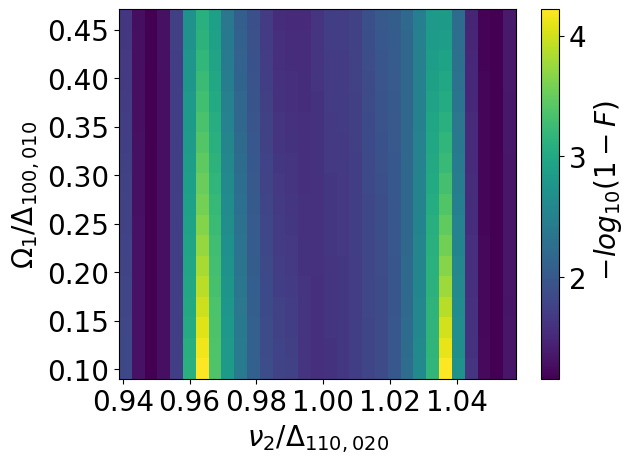}\put(-200,150){\textbf {(a)}}
\hfill
\includegraphics[width=0.4\textwidth]{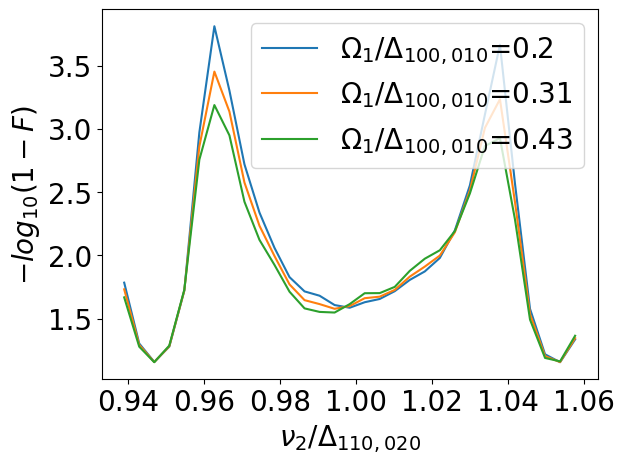}\put(-200,150){\textbf {(b)}}
 
\caption{Fidelity of the parametric cfSim gates with non-RWA terms included but no pulse-shaping. \ref{fig:non RWA fide}(a) Fidelity as a function of drive amplitude $\Omega_1$ and drive frequency$\nu_2$. Fidelity drop is observed in the middle area where drive amplitudes of the second drive $\Omega_2$ are large. This is because when non-RWA terms are included, the optimal amplitude formula Eq.~(\ref{eq:fsim opt amp final eq}) becomes less accurate. Under the calculated amplitude, leakage from state $|110\rangle$ is not minimized. \ref{fig:non RWA fide}(b) Three specific line cuts along the $\nu_2$ axis.}
\label{fig:non RWA fide}
\end{figure}

\begin{figure}
\includegraphics[width=0.4\textwidth]{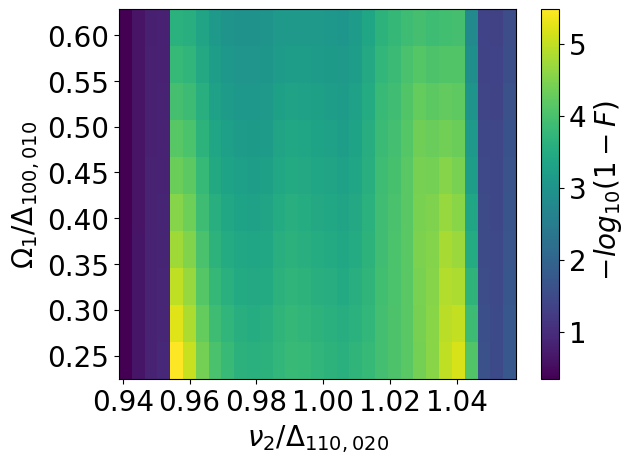}\put(-200,150){\textbf {(a)}}
\hfill
\includegraphics[width=0.4\textwidth]{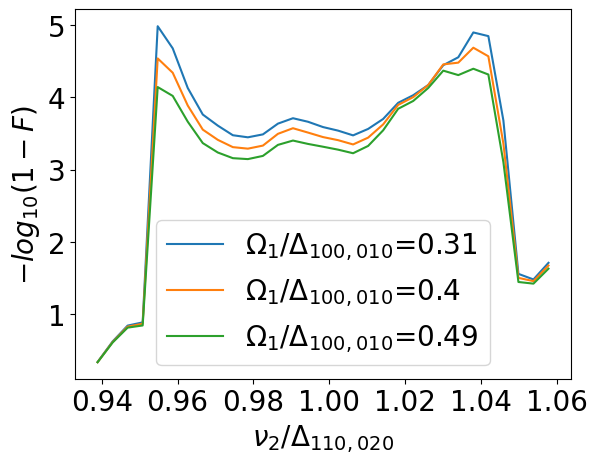}\put(-200,160){\textbf {(b)}}
 
\caption{Fidelity of the parametric cfSim gates with non-RWA terms included and pulse-shaping. \ref{fig:non RWA fide PS}(a) Fidelity as a function of drive amplitude $\Omega_1$ and drive frequency$\nu_2$. Fidelity is recovered after introducing pulse-shaping and numerically optimized amplitude $\Gamma$ in Eq.~(\ref{eq: op factor}). \ref{fig:non RWA fide PS}(b) Three specific line cuts along the $\nu_2$ axis.}
\label{fig:non RWA fide PS}
\end{figure}

We perform the same cfSim simulation as in Fig.~(\ref{fig:fSim num ana}) and (\ref{fig:fSim num fide log and linecut}). We find that in the middle area where the drives are strong, gate fidelity drops to around 99\%, see Fig.~(\ref{fig:non RWA fide}). This is due to the extra non-RWA terms. We have used the same amplitude formula in Eq.~(\ref{eq:fsim opt amp final eq}), which is analytically derived from the RWA Hamiltonian. In the middle area, where $\nu_2$ is close to $\Delta_{110,020}$, the required drive amplitude $\Omega_2$ is large. This gives a stronger non-RWA effect, resulting in the fidelity drop. Luckily, much of this non-RWA error can be corrected via pulse-shaping. We again use the same pulse-shaping and optimization procedure as in Eq.~(\ref{eq: op factor}) and perform the same simulation as in Fig.~(\ref{fig:fsim PS fid linecut}). The fidelity results are shown in Fig.~(\ref{fig:non RWA fide PS}). The fidelity loss is recovered after introducing a smooth envelope and using numerically optimized amplitudes. The non-RWA effects also change the iSWAP angle pattern while the conditional phases remain mostly unchanged. In general, iSWAP rate is reduced due to the non-RWA terms. Simulation results are included in Appendix (\ref{Appen:non RWA iswap cphase}).

Apart from non-RWA, another major nonlinear effect is the higher harmonics from drives\cite{roth2017analysis}. We have assumed the drive Hamiltonian takes the simple form of $H_d=\sum_{j=1,2} \Omega_j \sin(\nu_j t + \phi^d_j) \adag_j \ah_j$. A harmonic drive in the full circuit Hamiltonian Eq.~(\ref{eq:full circuit H in n p}) would generate higher-order tones, especially when the drive is strong. In theory, this nonlinearity can be reversed by pulse-shaping, too. Suppose there is a relation between the drive in the Fock space (Eq.~(\ref{eq:setup})) and the drive in the full circuit Hamiltonian Eq.~(\ref{eq:full circuit H in n p}), one can reverse engineer the drive envelope so that the drive in Eq.~(\ref{eq:setup}) is single-tone. This, of course, can be challenging to implement in experiments because the relation between the two is unknown in general. As an example, if the relation between two drives is given by $\omega(t)=\omega_0\sqrt{\varphi(t)/2}$\cite{PhysRevA.76.042319,DiCarlo_2009,roth2017analysis}, one can substitute in $\omega(t)=\Omega \sin (\nu t)$ to get $ \varphi(t)=2\left(\frac{\Omega}{\omega_0}\right)^2\sin^2(\nu t)$. This would then give the desired single-tone drive in Eq.~(\ref{eq:setup}).

Dealing with the strong drive effects is crucial for parametric drives. This is because, compared to gates without using parametric modulations, parametric gates are usually slower under the same hardware conditions. One solution is to find novel schemes to increase parametric gate speed without entering the strong drive regime. For instance, there has been a promising attempt at using extra levels to speed up parametric transitions from Chalmers University of Technology\cite{2025APS1,2025APS2}.

For future work, to get an accurate comparison with experiments and fully assess gate fidelities, full circuit Hamiltonians and open system simulations will be needed. This would help to understand what is the optimal gate time under trade-off between unitary errors and decoherence. Another direction is to apply the same simultaneous-driving principles to other gate setups. One of the easiest adaptations is to apply a parametric drive on top of a pair of resonant qubits so that iSWAP transition is activated by the resonance condition and CPHASE operation is controlled via the parametric drive. We provide some preliminary simulations in Appendix.
\ref{Appen:static + p fsim}.

\section{Summary}\label{sec.Summ}

In summary, we presented a proposal of implementing the continuous fSim gate in superconducting systems using bichromatic parametric drives. We propose to drive the iSWAP transition and CPHASE transition simultaneously. In order to study the crosstalk effect between the two drives, we derived accurate analytical formulas of the effective coupling strength valid for a wide range: from weak coupling to strong coupling and from weak drive to strong drive. With the help of our analytical formulas, we can calculate the optimal drive amplitude for minimizing leakage. The analytical calculation shows both iSWAP angle $\theta$ and conditional phase $\varphi$ can be tunned in full range, which is verified by numerical simulation. The flexibility of continuous fSim gates makes it advantageous to many quantum algorithms such as for simulating fermionic systems. The gate protocol is applicable to current superconducting quantum computation platforms. We hope it can facilitate the practical application of quantum computation in the NISQ era.

\begin{acknowledgments}

The authors thank Rami Barends, David DiVincenzo and Xuexin Xu for fruitful discussions and insightful comments. 
We acknowledge support from the Federal Ministry of Education and Research (BMBF) within the framework program ``Quantum technologies – from basic research to
market'', under the QSolid Project, Grant No. 13N16149.

\end{acknowledgments}


\bibliography{apssamp}

\clearpage

\appendix

\section{ Rotating Frame of the Toy Model}\label{Appen: rotating frame}

The time-dependent Hamiltonian $H(t)$ is given by:

\begin{widetext}
    \beq
\label{eq.APP Hqutrits}
    H_0=\begin{pmatrix}
  \Omega_2\sin(\nu_2 t) & g & 0 & 0 & 0 \\ 
  g & \Delta+\Omega_1\sin(\nu_1 t) & 0 & 0 & 0  \\
  0 & 0 & \Delta+\Omega_1\sin(\nu_1 t)+\Omega_2\sin(\nu_2 t) & \sqrt{2} g & \sqrt{2} g \\
  0 & 0 & \sqrt{2} g &  \delta+2\Omega_2\sin(\nu_2 t) & 0 \\
  0 & 0 & \sqrt{2} g & 0 & 2\Delta+\delta+2\Omega_1\sin(\nu_1 t) 
  \end{pmatrix}
\eeq

We collect all diagonal elements into $H_{\rm diag}$ and transform the Hamiltonian by the unitary matrix  $U=\exp(-i \int H_{\rm diag} (t) dt)$ into the interaction picture  --- i.e. $H_I= U^\dag H U-iU^\dag\frac{\partial}{\partial t} U$; equivalently using a rotating frame that is resonant with all levels---. The transformation matrix can be found as:

    \beq
\label{eq.APP Hqutrits}
    U=\begin{pmatrix}
  e^{i\frac{\Omega_2}{\nu_2}\cos(\nu_2 t)} & 0 & 0 & 0 & 0 \\ 
  0 & e^{i[\frac{\Omega_1}{\nu_1}\cos(\nu_1 t)-\Delta t]} & 0 & 0 & 0  \\
  0 & 0 & e^{i[\frac{\Omega_1}{\nu_1}\cos(\nu_1 t)+\frac{\Omega_2}{\nu_2}\cos(\nu_2 t)-\Delta t]} & 0 & 0 \\
  0 & 0 & 0 &  e^{i[\frac{2\Omega_2}{\nu_2}\cos(\nu_2 t)-\delta t]} & 0 \\
  0 & 0 & 0 & 0 & e^{i[\frac{2\Omega_1}{\nu_1}\cos(\nu_1 t)-2\Delta t-\delta t]} 
  \end{pmatrix}
\eeq

The transformed Hamiltonian $H_I$ can be easily calculated:

\beq
\label{eq.HIqutrits}
    H_I=g\begin{pmatrix}
  0 & e^{i[\frac{\Omega_1}{\nu_1}\cos(\nu_1 t)-\frac{\Omega_2}{\nu_2}\cos(\nu_2 t)-\Delta t]} & 0 & 0 & 0 \\ 
  h.c. & 0 & 0 & 0 & 0  \\
  0 & 0 & 0 & \sqrt{2}e^{i[\frac{\Omega_2}{\nu_2}\cos(\nu_2 t)-\frac{\Omega_1}{\nu_1}\cos(\nu_1 t)+\Delta t -\delta t]}  & \sqrt{2}e^{i[\frac{\Omega_1}{\nu_1}\cos(\nu_1 t)-\frac{\Omega_2}{\nu_2}\cos(\nu_2 t)-\Delta t-\delta t]}  \\
  0 & 0 & h.c.  &  0 & 0 \\
  0 & 0 & h.c.  & 0 & 0
  \end{pmatrix}
\eeq

where we have used $h.c.$ for Hermitian conjugate.

\end{widetext}

Simplifying Eq.(\ref{eq.HIqutrits}) using the Jacobi–Anger identity (i.e. $\exp({i \alpha \cos \theta})=\sum_n i^n J_n(\alpha) \exp({i n \theta})$, one gets the result Eq.(\ref{eq. HBessel}) in main text.

\section{Perturbative treatment of the toy model}\label{Appen:perutbation of toy model}

Without loss of generality, let us assume that the frequency of Q1 is greater than that of Q2, i.e. $\omega_1 > \omega_2$.  For this setup, we can parametrically drive the frequency of Q1 by supplying the required energy to achieve the transition $|01\rangle \leftrightarrow |10\rangle$, i.e. $\nu_1=\Delta_0$. The second qubit Q2 is parametrically driven to come very close to the resonance between $|11\rangle \leftrightarrow |02\rangle$ transition, i.e. $\nu_2=\Delta_0+\delta+\varepsilon$ with small $|\varepsilon|$, i.e. $ | \varepsilon | \ll \min\{ | \Delta_0 +\delta |, | \Delta_0 -\delta |\}$.  By this assumption, only the slow-varying terms proportional to $\exp(i \varepsilon t)$ contribute to the state evolution in the long run $t\gg 1/(\Delta_0\pm \delta)$.  The Hamiltonian (\ref{eq. HBessel}) contains interaction within and outside the Hilbert space's computational subset. Within the computational subset there is exchange interaction $ |10\rangle \langle 01|$ with the strength  $gJ_1(\Omega_1/\Delta_0) J_0 (\Omega_2/(\Delta_0+\delta+\varepsilon))$. Outside of the computational subset the energy level $|11\rangle$ interacts with $|02\rangle$ and $|20\rangle$ by the following coupling strength $\sqrt{2}g J_0 (\Omega_1/\Delta_0) J_1(\Omega_2/(\Delta_0+\delta+\varepsilon)) e^{i\varepsilon t}$  and $\sqrt{2}g J_{-2} (\Omega_1/\Delta_0) J_1(\Omega_2/(\Delta_0+\delta+\varepsilon)) e^{i\varepsilon t}$, respectively.

One of the aspects of perturbative validity of our result is that the qubits are weakly driven, therefore we consider that the drive amplitude is weak, i.e. ${\Omega_i}<\nu_i$ for $i=1,2$. Within this limit which makes the arguments of the Bessel function, small, one can show that $J_{-2} J_1 \ll J_1 J_0$; therefore, the coupling strength of $|11\rangle \langle 20|$ transition turns out to be negligible compared to the other two interaction strengths and can be approximately ignored.  Therefore in the interaction picture after rotating-wave-approximation, the BPD Hamiltonian can be simplified as 
\begin{eqnarray}
\label{eq. HBessel_simplf}
      H  &=& gJ_1(x_1) J_0(x_2) |10\rangle \langle 01| \nonumber \\ && + \sqrt{2} gJ_0(x_1) J_1(x_2) e^{i\varepsilon t}|11\rangle \langle 02| + h.c.
\end{eqnarray}

\section{Analytical coupling strength by exact diagonalization}\label{appen:G}

For more accuracy, one can numerically find the exact transformation to diagonalize the undriven Hamiltonian.

Assuming we can find the exact transformation $U^{exa}$ for the undriven Hamiltonian $H$:

\[U=\sum u_{Q_1^\prime Q_2^\prime R^\prime, Q_1 Q_2 R} |Q_1^\prime Q_2^\prime R^\prime\rangle \langle Q_1 Q_2 R|\]

The undriven Hamiltonian is diagonalised by this transformation:

\[\Tilde{H}=U^{\dag} H U =\sum_{Q_1 Q_2 R} \Tilde{\omega} _{Q_1 Q_2 R} |Q_1 Q_2 R\rangle \langle Q_1 Q_2 R|\]

The drive Hamiltonian can be transformed to the diagonal frame by the same transformation $U$:

\begin{equation} \label{eq2_new}
\begin{split}
\Tilde{H}_{d}&=U^{\dag} H_d U\\
&=\sum_i \Omega_i \sin(\nu_i t+\phi_i^d)U^{\dag} a^\dag_ia_i U\
\end{split}
\end{equation}

$U^{\dag} a^\dag_ia_i U$ can be numerically evaluated as well:

\begin{widetext}
\begin{equation} \label{eq: appen U QQR}
\begin{split}
U^{\dag} a^\dag_m a_m U&=\sum  |Q_1^\prime Q_2^\prime R^\prime\rangle \langle Q_1^\prime Q_2^\prime R^\prime|U^{\dag} a^\dag_m a_m U |Q_1 Q_2 R\rangle \langle Q_1 Q_2 R|\\
&=\sum\langle Q_1^\prime Q_2^\prime R^\prime|U^{\dag} a^\dag_m a_m U |Q_1 Q_2 R\rangle |Q_1^\prime Q_2^\prime R^\prime\rangle  \langle Q_1 Q_2 R|
\end{split}
\end{equation}

Using the notations introduced in Eq.(\ref{eq:N C}), we can split Eq.(\ref{eq: appen U QQR}) further into diagonal terms, which correspond to correction to energies, and off-diagonal terms, which couples different eigenstates of the undriven Hamiltonian:

\begin{equation} \label{eq3}
U^{\dag} a^\dag_m a_m U=\sum_{Q_1 Q_2 R}  N^m_{Q_1 Q_2 R} |Q_1 Q_2 R\rangle \langle Q_1 Q_2 R|+\sum_{Q_1^\prime Q_2^\prime R^\prime \neq Q_1 Q_2 R}  C^m_{Q_1^\prime Q_2^\prime R^\prime,Q_1 Q_2 R} |Q_1^\prime Q_2^\prime R^\prime\rangle \langle Q_1 Q_2 R|
\end{equation}

Substituting Eq.(\ref{eq3}) into Eq.(\ref{eq2_new}) we get:

\begin{equation} \label{eq4}
\begin{split}
\Tilde{H}_{d}&=U^{\dag} H_d U\\
=&\sum \Omega_m \sin(\nu_m t+\phi^d_m) (N^m_{Q_1 Q_2 R} |Q_1 Q_2 R\rangle \langle Q_1 Q_2 R|+C^m_{Q_1^\prime Q_2^\prime R^\prime,Q_1 Q_2 R} |Q_1^\prime Q_2^\prime R^\prime\rangle \langle Q_1 Q_2 R|)\\
=&\sum \Omega_m \sin(\nu_m t+\phi^d_m)  N^m_{Q_1 Q_2 R} |Q_1 Q_2 R\rangle \langle Q_1 Q_2 R|+\sum \Omega_m \sin(\nu_m t+\phi^d_m) C^m_{Q_1^\prime Q_2^\prime R^\prime,Q_1 Q_2 R} |Q_1^\prime Q_2^\prime R^\prime\rangle \langle Q_1 Q_2 R|\
\end{split}
\end{equation}

The total Hamiltonian now reads as follows:

\begin{equation} \label{eq:appen total H diag frame}
\begin{split}
\Tilde{H}&=\sum_{Q_1 Q_2 R} (\Tilde{\omega} _{Q_1Q_2R}+ \sum_m \Omega_m \sin(\nu_m t+\phi^d_m)  N^m_{Q_1Q_2R} )|Q_1Q_2R\rangle \langle Q_1Q_2R|\\
&+\sum_{m Q_1 Q_2 R \neq Q_1^\prime Q_2^\prime R^\prime} \Omega_m \sin(\nu_m t+\phi^d_m) C^m_{Q_1^\prime Q_2^\prime R^\prime, Q_1 Q_2 R} |Q_1^\prime Q_2^\prime R^\prime\rangle \langle Q_1 Q_2 R|\
\end{split}
\end{equation}

The rotating frame is now defined by the transformation:

\begin{equation} \label{eq6}
\begin{split}
U_r&=e^{-i \int \sum_{Q_1 Q_2 R} (\Tilde{\omega} _{Q_1Q_2R}+ \sum_m \Omega_m \sin(\nu_m t+\phi^d_m)  N^m_{Q_1Q_2R} )|Q_1Q_2R\rangle \langle Q_1Q_2R| dt}\\
&=\sum_{Q_1 Q_2 R} e^{-i \int  ( \Tilde{\omega} _{Q_1Q_2R}+ \sum_{m}\Omega_m \sin(\nu_m t+\phi^d_m)  N^m_{Q_1Q_2R} ) dt}|Q_1Q_2R\rangle \langle Q_1Q_2R|
\end{split}
\end{equation}

The final Hamiltonian in the rotating frame can be written as:

\begin{equation}\label{eq7}
\Tilde{H}_{R}=\sum G(t)_{Q_1^\prime Q_2^\prime R^\prime, Q_1 Q_2 R} |Q_1^\prime Q_2^\prime R^\prime\rangle \langle Q_1 Q_2 R|
\end{equation}

where

    \begin{equation}\label{eq8}
\begin{split}
G(t)_{Q_1^\prime Q_2^\prime R^\prime, Q_1 Q_2 R} &=  \sum_{n_r, n_1, n_2=-\infty}^{+\infty} \sum_{m=r,1,2}A_{n_r n_1 n_2}^{Q_1^\prime Q_2^\prime R^\prime, Q_1 Q_2 R}(m)\Omega_m C^m_{Q_1^\prime Q_2^\prime R^\prime, Q_1 Q_2 R} i^{n_r+n_1+n_2} \\
&*e^{i((\Tilde{\omega} _{Q_1^\prime Q_2^\prime R^\prime}-\Tilde{\omega} _{Q_1 Q_2 R}-n_r \nu_r-n_1 \nu_1-n_2 \nu_2 )t-n_r \varphi_r-n_1 \varphi_1- n_2 \varphi_2) }   
\end{split}
\end{equation}

\begin{equation} \label{eq9}
\begin{split}
A_{n_r n_1 n_2}^{Q_1^\prime Q_2^\prime R^\prime, Q_1 Q_2 R}(r)=&\frac{J_{n_r+1}(-s_r^{Q_1^\prime Q_2^\prime R^\prime, Q_1 Q_2 R})+J_{n_r-1}(-s_r^{Q_1^\prime Q_2^\prime R^\prime, Q_1 Q_2 R})}{2}\\
&*J_{n_1}(-s_1^{Q_1^\prime Q_2^\prime R^\prime, Q_1 Q_2 R})J_{n_2}(-s_2^{(ijk,npq)})\\
A_{n_r n_1 n_2}^{Q_1^\prime Q_2^\prime R^\prime, Q_1 Q_2 R}(1)=&J_{n_r}(-s_r^{Q_1^\prime Q_2^\prime R^\prime, Q_1 Q_2 R})\\
&*\frac{J_{n_1+1}(-s_1^{Q_1^\prime Q_2^\prime R^\prime, Q_1 Q_2 R})+J_{n_1-1}(-s_1^{Q_1^\prime Q_2^\prime R^\prime, Q_1 Q_2 R})}{2}\\
&*J_{n_2}(-s_2^{Q_1^\prime Q_2^\prime R^\prime, Q_1 Q_2 R})\\
A_{n_r n_1 n_2}^{Q_1^\prime Q_2^\prime R^\prime, Q_1 Q_2 R}(2)=&J_{n_r}(-s_0^{Q_1^\prime Q_2^\prime R^\prime, Q_1 Q_2 R})J_{n_1}(-s_1^{Q_1^\prime Q_2^\prime R^\prime, Q_1 Q_2 R})\\
&*\frac{J_{n_2+1}(-s_2^{Q_1^\prime Q_2^\prime R^\prime, Q_1 Q_2 R})+J_{n_2-1}(-s_2^{Q_1^\prime Q_2^\prime R^\prime, Q_1 Q_2 R})}{2}
\end{split}
\end{equation}
\end{widetext}

\[s_m^{Q_1^\prime Q_2^\prime R^\prime, Q_1 Q_2 R}=\frac{N^m_{Q_1^\prime Q_2^\prime R^\prime}-N^m_{Q_1 Q_2 R}}{\nu_m}\Omega_m\]

Note that here the definition of coupling strength $G(t)$ is different from usual definition of $g$ by the factor $\sqrt{(n_j+1)n_k}$.

The resonance condition is now given by:

\begin{equation}\label{eq:res exact2}
\Tilde{\omega} _{Q_1^\prime Q_2^\prime R^\prime}-\Tilde{\omega} _{Q_1 Q_2 R}-n_r \nu_r-n_1 \nu_1-n_2 \nu_2 =0
\end{equation}

\section{ Non-RWA effects on iSWAP angles and conditional phases }\label{Appen:non RWA iswap cphase}

\begin{figure*}[!ht]
\includegraphics[width=0.35\textwidth]{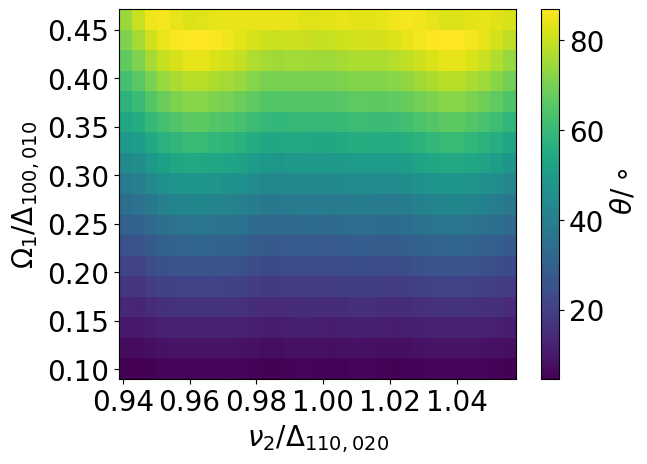}\put(-170,130){\textbf {(a)}}
\includegraphics[width=0.35\textwidth]{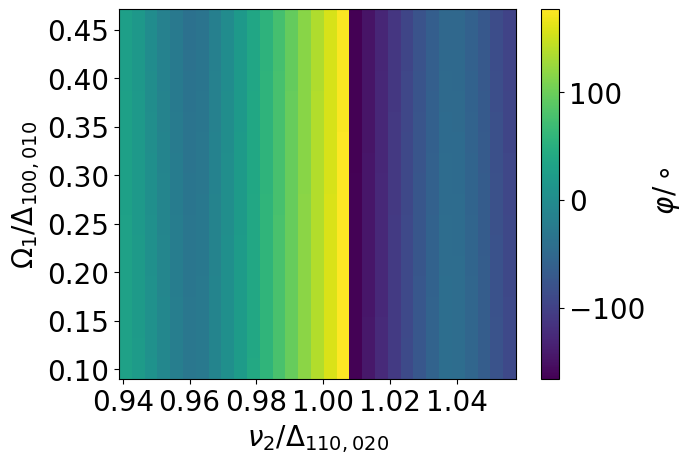}\put(-170,130){\textbf {(b)}}\\
\includegraphics[width=0.35\textwidth]{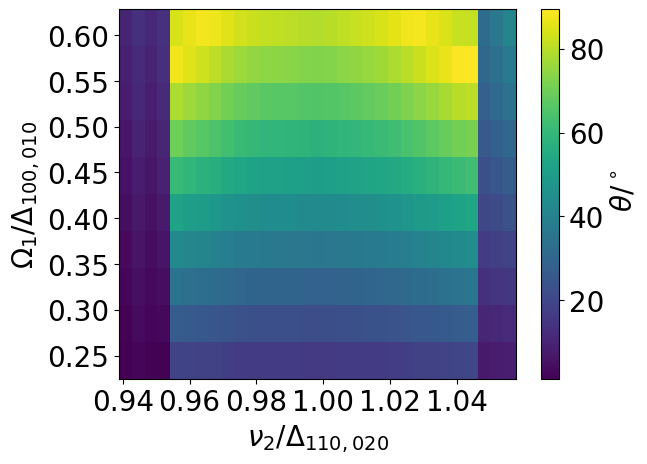}\put(-170,130){\textbf {(c)}}
\includegraphics[width=0.35\textwidth]{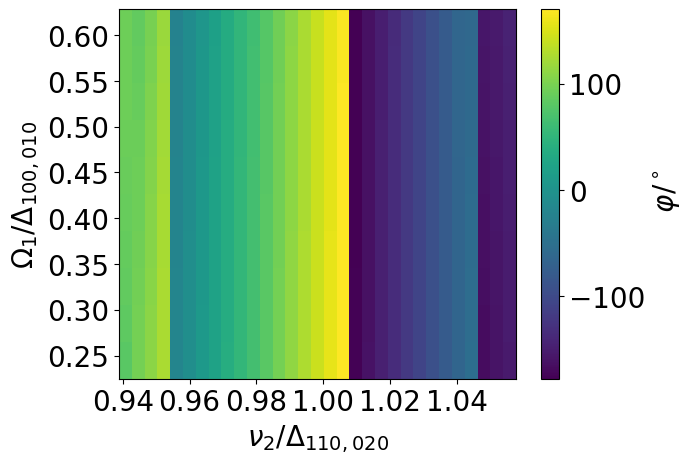}\put(-170,130){\textbf {(d)}}
        
\caption{iSWAP and CPHASE patterns with non-RWA terms included. \ref{fig:non RWA iswap cphase}(a) iSWAP pattern without pulse-shaping. The iSWAP pattern is shifted towards larger $\Omega_1$. This is likely because the effective coupling between $|100\rangle$ and $|010\rangle$ is reduced by the non-RWA terms. \ref{fig:non RWA iswap cphase}(b) CPHASE pattern without pulse-shaping. The CPHASE pattern remains mostly unchanged. \ref{fig:non RWA iswap cphase}(c) iSWAP pattern with pulse-shaping. \ref{fig:non RWA iswap cphase}(d) CPHASE pattern with pulse-shaping. Similar as the patterns without pulse-shaping.}
\label{fig:non RWA iswap cphase}
\end{figure*}

The iSWAP angles and conditional phases are not so much affected by non-RWA effects as fidelity. We show the simulation results in FIG.~(\ref{fig:non RWA iswap cphase}).

\section{ fSim gates via DC mixed with AC drive.}\label{Appen:static + p fsim}

The simultaneous transition scheme is very flexible. Instead of using two parametric drives for two transitions, one can activate one transition via a DC drive ( two levels tuned on resonance) and drive the other transition via a parametric drive. Here we use the two-qutrit toy model to demonstrate the idea. 

We drive the iSWAP transition by tuning $|01\rangle$ and $|10\rangle$ into resonance and drive the CPHASE transition via a parametric drive. For simplicity, let's assume we can directly control the coupling strength $g(t)=g_0+\Omega \sin(\omega t)$. The time-dependent coupling strength between two qutrits has a DC part $g_0$ and a parametrically driven part $\Omega \sin(\omega t)$. Similar as Eq.(\ref{eq.Hqutrits}), we can then write down the Hamiltonian in the Hilbert space spanned by $\{ |01\rangle, |10\rangle, |11\rangle, |02\rangle, |20\rangle\}$:

\beq
\label{eq.Hqutrits DC AC}
    H(t)=\begin{pmatrix}
  0 & g(t) & 0 & 0 & 0 \\ 
  g(t) & 0 & 0 & 0 & 0  \\
  0 & 0 & 0 & \sqrt{2} g(t) & \sqrt{2} g(t) \\
  0 & 0 & \sqrt{2} g(t) &  \delta_2 & 0 \\
  0 & 0 & \sqrt{2} g(t) & 0 & \delta_1 
  \end{pmatrix}
\eeq

\begin{figure*}[h!]
\centering
\begin{subfigure}{0.32\linewidth}
    \includegraphics[width=\linewidth]{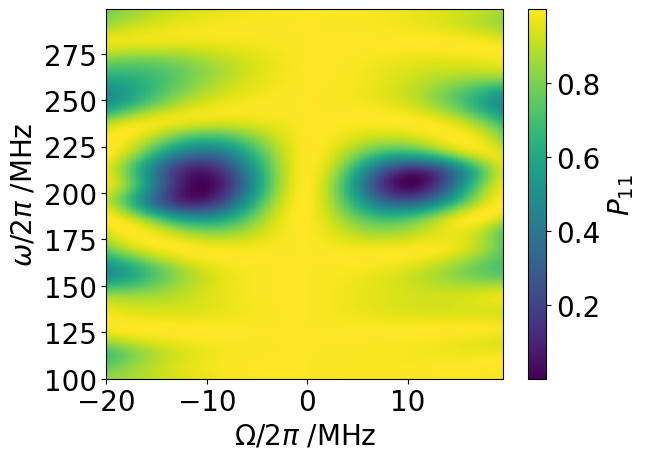}
    \put(-155,120){\textbf{(a)}}
\end{subfigure}
\begin{subfigure}{0.32\linewidth}
    \includegraphics[width=\linewidth]{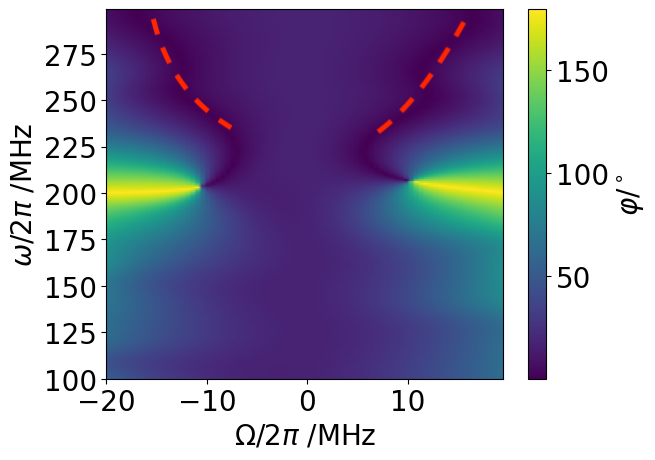}\put(-155,120){\textbf{(b)}}
\end{subfigure}
\begin{subfigure}{0.32\linewidth}
    \includegraphics[width=\linewidth]{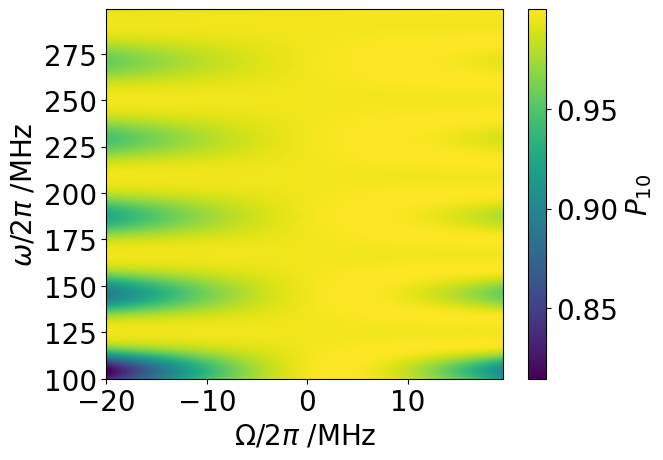}\put(-155,120){\textbf{(c)}}
\end{subfigure}
\caption{Numerical simulation of DC-AC driven fSim gates. We assume two qutrits have same anharmonicity $\delta_1/2\pi=\delta_2/2\pi= -200$ MHz. Gate time $t_g$ is chosen to be around 25 ns. The DC drive $g_0/2\pi$ is $10$ MHz. \ref{fig: stat para}(a)Population of $|11\rangle$. The two leakage centers occur when AC drive frequency $\omega$ is matching with anharmonicity causing resonant oscillations among $|11\rangle$, $|02\rangle$ and $|20\rangle$. \ref{fig: stat para}(b) Conditinal phase of $|11\rangle$. A $\pi$ phase can be realized under resonant AC drive, corresponding to a CZ operation. The zero conditional phase trajectories are marked with red dashed lines. \ref{fig: stat para}(c) Population of $|11\rangle$. Similar as the BPD case, there is some crosstalk from the AC drive to the iSWAP rate. This can be understood as baseband modulation.}
\label{fig: stat para}
\end{figure*}

We numerically solve the time-depedent Schr\"{o}dinger equation under this Hamiltonian. Similar behaviours as BPD dynamics can be observed. Given a fixed gate time, there are "leakage centers" corresponding to population of $|11\rangle$ being coherently driven out of the computational space. Away from those centers, leakage is small. The conditional phase is tunable by varying drive amplitude and frequency. iSWAP rate is mainly controlled by the DC drive $g_0$. In the simulation, we have chosen gate time $t_g$ and DC drive $g_0$ so that iSWAP angle $\theta$ is around $90 ^\circ$, see Fig. (\ref{fig: stat para})

\end{document}